\documentclass[12pt,a4paper,twoside,final,notitlepage,reqno]{article}
\usepackage[english]{babel}
\usepackage[T1]{fontenc}
\usepackage{epsfig, graphicx, amssymb}
\usepackage{amsmath,amsthm,epsfig,amsfonts}
\usepackage{float}
\usepackage{color}
\newcommand{\moneq}{\vspace*{-7pt} \begin{equation} \displaystyle }
\newcommand{\moneqstar}{\vspace*{-6pt} \begin{equation*} \displaystyle }
\newcommand{\monendstar}{\vspace*{-6pt} \end{equation*}   }
\newcommand{\monend}{\vspace*{-7pt} \end{equation}   }
\newcommand{\moneqarraystar}{ \begin{eqnarray*} \displaystyle }
\newcommand{\monendarraystar}{ \end{eqnarray*}   }

\newcommand{\dd}{{\rm d}}

\definecolor{vertfonce}{rgb}{0.0, 0.5, 0.0}

\def\section*#1{}

\usepackage{fancyhdr}
\renewcommand{\headrulewidth}{0pt}

\begin{document}

\fancypagestyle{plain}{ \fancyfoot{} \renewcommand{\footrulewidth}{0pt}}
\fancypagestyle{plain}{ \fancyhead{} \renewcommand{\headrulewidth}{0pt}}

~

  \vskip 2.1 cm

\centerline {\bf \LARGE Multiresolution relaxation times}

 \bigskip 

\centerline {\bf \LARGE lattice Boltzmann schemes with projection}

\bigskip  \bigskip \bigskip

\centerline { \large   Fran\c{c}ois Dubois$^{ab}$ and Paulo Cesar Philippi$^{c}$}

\smallskip  \bigskip

\centerline { \it  \small
  $^a$ Laboratoire de Math\'ematiques d'Orsay, Facult\'e des Sciences d'Orsay,}

\centerline { \it  \small   Universit\'e Paris-Saclay, France.}

\centerline { \it  \small
$^b$    Conservatoire National des Arts et M\'etiers, LMSSC laboratory,  Paris, France.}

\centerline { \it  \small
  $^c$   Pontifícia Universidade Cat\'olica do Paran\'a,  Curitiba, Paran\'a, Brazil.}



\bigskip  \bigskip

\centerline {20  December 2024 
{\footnote {\rm  \small $\,$ This contribution, presented 
at the 33rd Conference on Discrete Simulation of Fluid Dynamics,
Eidgen\"ossische Technische Hochschule 
Z\"urich (Switzerland) the 09 July 2024,
is published in the journal {\it Physics of  Fluids}, volume 37, issue 03, article 037179,  2025.
}}}




\bigskip  \bigskip
\noindent {\bf \large Abstract}

\noindent
We propose to extend the multiresolution relaxation times lattice Boltzmann schemes with
a additional  projection step. For the explicit example of the D2Q9 scheme, we define this
extended method. We prove that in general the projection step does not change the asymptotic
partial differential equations at second order.
We present four numerical test cases. One concerns linear stability
with a Fourier analysis with a single-vertex scheme. Three bidimensional fluid flows
with a coarse mesh have been tested:
the Minion and Brown 
sheared flow,  the Ghia, Ghia and Shin 
lid-driven cavity and an unsteady acoustic wave.
Our results indicate that the 
bulk viscosity can be dramatically reduced with a better stability than the initial scheme.

\noindent

\newpage

\bigskip \bigskip    \noindent {\bf \large    1) \quad  Introduction}

\fancyhead[EC]{\sc{Fran\c{c}ois Dubois and Paulo Cesar Philippi}}
\fancyhead[OC]{\sc{MRT lattice Boltzmann schemes with projection}}
\fancyfoot[C]{\oldstylenums{\thepage}}

\smallskip \noindent
The single relaxation time lattice Boltzmann schemes proposed by
Higuera and Jim\'enez~\cite {HJ89},
McNamara and Zanetti \cite{MZ88}, 
Qian, d'Humi\`eres and  Lallemand \cite{QHL92}
have two origins: the  lattice gas automata
of Hardy, Pomeau and de Pazzis \cite{HPP73}
and Frisch, Hasslacher and Pomeau~\cite{FHP86}, 
and the  discrete velocities models for the Boltzmann equation
introduced by Carleman \cite{Ca57}, Broadwell \cite{Br64}, 
and Gatignol \cite{Ga75}.

\smallskip \noindent
The  multiresolution relaxation times (MRT) lattice Boltzmann scheme
is essentially due to d'Humi\`eres \cite{DDH92}.
Two representations of the discrete gas are used: a particle representation
and a  representation of the state  with  set of moments.
The macroscopic moments are conserved during the collision 
and satisfy asymptotically  a set of macroscopic
partial differential equations.
The non-equilibrium moments   
are not conserved and relax towards an equilibrium function
using several  relaxation times.
This method is well understood since the work of Lallemand and Luo \cite{LL00}.

\smallskip \noindent
In a remarkable 1994 paper, Ladd \cite{La94} introduced two new ingredients in writing athermal
second-order LB equations. First, the collision operator was linearized giving rise to a kinetic
model written in terms of second-order moments, with two independent collision frequencies.
Second, the relaxation of the viscous stress tensor was decomposed into the relaxation of its
deviatoric and isotropic parts.

\smallskip \noindent
A new idea has been proposed by  Shan {\it et al.} \cite{SYC06} 
and Philippi  {\it et al.} \cite{PHDS06} with the introduction
of Hermite polynomials to represent the moments at equilibrium. A simplification of the method,
so-called ``regularization'' has been proposed by  Latt and Chopard \cite{LC06}.
The  recursivity and regularization  of  Malaspinas  \cite{Ma15} is an extension of Latt's algorithm.
The high-order regularization of Mattila {\it et al.} \cite{MPH17} is the basis of developments
done by one of us. 
The initial MRT algorithm is simplified and the number of parameters is reduced.
But it is necessary to have a good knowledge of the algebraic properties
of the Hermite polynomials and their links with transport properties of fluids.

\smallskip \noindent
In this contribution, we follow a totally discrete approach without any need of Hermite polynomials.
The Taylor expansion method \cite{Du08} has been extended with the introduction of the
differential advection operator in the basis of moments and the ``ABCD'' decomposition~\cite{Du22}.
Then the analysis of MRT lattice Boltzmann schemes can be conducted without any {\it a priori} reference
to fluid properties. The equivalent partial differential equations can be  formally  derived from
the knowledge of the equilibrium value of the non-equilibrium  moments and the relaxation parameters.  

\smallskip \noindent
We have observed in \cite{DL23} that three families of moments emerge from
the asymptotic analysis of lattice Boltzmann schemes: 
the conserved  moments  $ W $ that define the unknowns of the equivalent partial differential equations, 
the nonconserved    ``viscous  moments'' $ Y_e $ for setting the first order terms
and the nonconserved    ``energy transfer moments''  $ Y_v $ for adjusting second-order dissipation.
Precise definitions of these quantities are given below. 

\smallskip \noindent
In this contribution, we propose a new ``multiresolution relaxation times with projection'' lattice Boltzmann
scheme  inspired by kinetic regularization involving  Hermite polynomials 
(see {\it e.g.} \cite{LC06, Ma15, MPH17} and many others!).
Our motivation is to be able to make the computations very near the stability limit.
In that case, the use of a coarse mesh is possible and the global cost of the
computation is reduced.

\smallskip \noindent
The  outline of our contribution is the following.
We remind in section 2 the essential about  the multiple relaxation time  schemes for fluids,
in particular for the D2Q9 lattice Boltzmann scheme.
Then in section 3, we explain how the 
equivalent partial equivalent differential equations emerge from an asymptotic analysis
based on the  ABCD  decomposition. 
The present projected   multiple relaxation time D2Q9 scheme is presented in section 4. 
We insist on the importance of the hollow structure of the matrix of advection in the basis of moments. 
The asymptotic analysis of the MRT scheme with projection is conducted in section 5. 
The first numerical experiments for a linear model problem are presented in section 6.
In section~7, we focus on two  fluid applications for two space dimensions.
An unsteady linear acoustics problem is presented in section 8.
Some words of conclusion are proposed in section 9.
The section 10 is an appendix developing a technical point relative to the 
multiresolution relaxation times lattice Boltzmann schemes with projection. 

\bigskip \bigskip    \noindent {\bf \large    2) \quad  {Multiple relaxation time D2Q9 scheme for fluids}}

\begin{figure}    [H]  \centering
\centerline{\includegraphics[width=.23\textwidth]{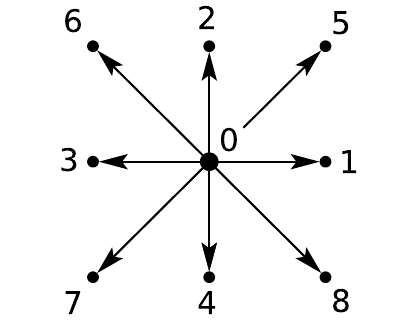}} 
\caption{The  nine velocities of the D2Q9 scheme  \cite{LL00}.}
\label{fig-d2q9} \end{figure}

\smallskip \noindent
In this section, we recall the basics about the  D2Q9 lattice Boltzmann scheme  for isothermal fluid flow,
studied in detail by Lallemand and Luo \cite{LL00}.
Recall that in the MRT framework proposed by d'Humi\`eres \cite{DDH92},
the mesoscopic scale is represented with the vector $ \, f \, $ 
describing the distribution of particle populations over the discrete set
  $ \, e_j \, $ for $ \, j = 0, \dots, b-1 \, $ of microscpîc velocities,  
and the vector $ \, m \, $ of moments.
The vector of particles is associated to a velocity rose
described on a square lattice at  the Figure \ref{fig-d2q9}.
A scale speed $ \, \lambda \,$ is associated to the ratio between the space step $ \, \Delta x \, $
and the time step $ \, \Delta t $:
\moneqstar 
 \lambda = {{\Delta x}\over{\Delta t}}  \, .
\monendstar 
%

\noindent
The velocities $ \, e_j \, $ for $ \, 0 \leq j \leq b \equiv 8 \, $ of the D2Q9 scheme described in Figure \ref{fig-d2q9}
admit the components $ \, e_{jx} , e_{jy} $. They 
are scaled with the numerical velocity $ \, \lambda $. We have 
\moneqstar
\{  e_j \} = \begin{pmatrix}
0 & \lambda & 0 & -\lambda & 0 & \lambda & -\lambda &  -\lambda &  \lambda \\
0 & 0 &  \lambda & 0 & -\lambda &  \lambda &  \lambda &  -\lambda &  -\lambda \end{pmatrix} \, . 
\monendstar 
The  particles and the moments are   linked together with the ``d'Humi\`eres matrix'' $ \, M \, $
such that 
\moneq \label{f-to-m}
m = M \, f \, .
\monend
\smallskip \noindent  
Following \cite{LL00}, the matrix $ \, M \, $ is constructed with the help of polynomials
relative to the velocities $ \, e_i $. We have
\moneqstar
M_{ij} = P_i (e_{jx} , e_{jy}) 
\monendstar 
with a family of polynomials given  in \cite{LL00} by     
\moneqstar        
\left\{  \begin{array} {rcl}  
  P_0  (e_{jx} , e_{jy})  &=& 1 \\
  P_1  (e_{jx} , e_{jy}) &=& e_{jx} \\ 
  P_2  (e_{jx} , e_{jy}) &=& e_{jy} \\
  P_3  (e_{jx} , e_{jy}) &=& 3 \, ( e_{jx}^2 + e_{jy}^2 ) - 4 \, \lambda^2 \\
  P_4  (e_{jx} , e_{jy}) &=& e_{jx}^2 - e_{jy}^2  \\
  P_5  (e_{jx} , e_{jy}) &=& e_{jx} \, e_{jy}    \\
  P_6  (e_{jx} , e_{jy}) &=&  \big[ 3 \, ( e_{jx}^2 + e_{jy}^2 )  - 5 \, \lambda^2 \big] \, e_{jx}  \\
  P_7  (e_{jx} , e_{jy}) &=&  \big[ 3 \, ( e_{jx}^2 + e_{jy}^2 )  - 5 \, \lambda^2 \big] \, e_{jy}  \\
  P_8  (e_{jx} , e_{jy}) &=&  {9\over2} \, ( e_{jx}^2 + e_{jy}^2 )^2 - {21\over2}  \, ( e_{jx}^2 + e_{jy}^2 )\, \lambda^2 + 4 \, \lambda^4    
\end{array} \right.  \,.
\monendstar   
%

\noindent
The matrix $ \, M \, $ is an invertible fixed matrix. Following  \cite{LL00}, its lines  are chosen orthogonal and each line corresponds to a specific
moment. We have for the D2Q9 scheme: 
\moneq \label{M-d2q9}
M =  \left[  \!\!   \begin{array} {ccccccccc}
 1  \!&\!  1  \!&\!  1  \!&\!  1  \!&\!  1  \!&\!  1  \!&\!  1  \!&\!  1  \!&\!  1 \\ 
   0 \!&\! \lambda  \!&\!  0  \!&\!  -\lambda  \!&\!  0  \!&\!   \lambda  \!&\!   -\lambda   \!&\!  -\lambda  \!&\!  \lambda  \\ 
 0  \!&\!  0   \!&\! \lambda  \!&\!  0    \!&\!   -\lambda   \!&\!   \lambda   \!&\!   \lambda  \!&\!  -\lambda  \!&\!  -\lambda \\ 
 -4 \lambda^2    \!&\!  -\lambda^2   \!&\!   -\lambda^2   \!&\!   -\lambda^2   \!&\!   -\lambda^2  \!&\!   2 \lambda^2   \!&\!
 2 \lambda^2   \!&\!  2 \lambda^2   \!&\!  2 \lambda^2 \\ 
  0    \!&\!   \lambda^2    \!&\!  -\lambda^2    \!&\!   \lambda^2   \!&\!   -\lambda^2  \!&\!  0  \!&\!  0  \!&\!  0  \!&\!  0 \\ 
 0  \!&\!  0  \!&\!  0  \!&\!  0  \!&\!  0   \!&\!   \lambda^2   \!&\!   -\lambda^2    \!&\!   \lambda^2   \!&\!   -\lambda^2 \\ 
 0   \!&\! -2 \lambda^3  \!&\!  0   \!&\!  2 \lambda^3  \!&\!  0   \!&\!  \lambda^3   \!&\!   -\lambda^3   \!&\!   -\lambda^3   \!&\!    \lambda^3 \\
 0  \!&\!  0  \!&\! -2 \lambda^3  \!&\!  0  \!&\!  2 \lambda^3    \!&\!   \lambda^3    \!&\!   \lambda^3    \!&\!  -\lambda^3    \!&\!  -\lambda^3 \\ 
4 \lambda^4  \!&\! -2 \lambda^4  \!&\! -2 \lambda^4  \!&\! -2 \lambda^4  \!&\! -2 \lambda^4  \!&\!   \lambda^4   \!&\!  \lambda^4   \!&\!  \lambda^4  \!&\!   \lambda^4 
  \end{array} \!\! \right] \,.
\monend
In this contribution, the nine moments are represented with  the following notations: 
\moneqstar
m = \big( \rho ,\, J_x ,\, J_y  ,\, \varepsilon  ,\,  p_{xx}  ,\, p_{xy}  ,\, q_x   ,\, q_y  ,\, h \big)^{\rm t}  .
\monendstar 
We observe that we have by definition 
$ \, p_{xx} = \sum_j  \big( e_{jx}^2 - e_{jy}^2 \big) \, f_j \, $ and $ \, p_{xy} = \sum_j  e_{jx} \, e_{jy} \, f_j $.
Moreover, following Lallemand and Luo \cite{LL00},
the three moments  $ \, \varepsilon $, $ \,  p_{xx} \, $ and $ \,  p_{xy} \, $
associated with polynomials of degree two define  
 a second-order tensor that describes the transfer of
momentum, including the tensor $ \, \tau_{\alpha \, \beta} \, $ responsible for the viscous transfer,  
while $\, q_x \, $  and $ \, q_y \, $  are third-order moments describing the
transfer of energy.
%
%

\noindent
The density $ \, \rho \, $ and the two components $ \, (J_x,\, J_y) \, $ of the momentum
constitute the vector $ \,W \, $ of conserved variables:
\moneq \label{W-d2q9}
 W =  (\rho ,\, J_x ,\, J_y )^{\rm t} \, .
\monend
The other moments 
\moneq \label{Y-d2q9}
Y = ( \varepsilon  ,\,  p_{xx}  ,\, p_{xy}  ,\, q_x   ,\, q_y  ,\, h )^{\rm t} 
\monend
are the  non-equilibrium  moments.
We have
\moneq \label{m-WY-d2q9}
m = \begin{pmatrix} W \\ Y \end{pmatrix} \, .
\monend

\bigskip

The relaxation process constructs locally a new vector of moments denoted by $ \, m^* \, $
with a local and nonlinear algorithm.
First the conserved moments are invariant in this process:
\moneqstar 
\rho^* = \rho  ,\, J_x^* = J_x  ,\, J_y^* = J_y
\monendstar
and we have simply $ \,  W^* = W  $.
Secondly, we define a vector $\, \Phi(W) \, $ of non-conserved moments at equilibrium.
 We introduce the two components $\, (u,\, v) \, $ of the macroscopic velocity
with the relations 
\moneqstar
J_x \equiv \rho \ u \,,\,\,  J_y = \rho \, v .
\monendstar
For the D2Q9 scheme for fluid flows, we have the classical relations \cite{CD98,QHL92}:
%
\moneq \label {formules-Phi-d2q9}
\left\{ \begin{array} {rcl}
  \Phi_{\varepsilon}  &=&   -  2 \, \lambda^2 \, \rho \, + \, 3 \, \rho \, ( u^2 + v^2 )  \\
  \Phi_{xx} &=&   \rho \, ( u^2 - v^2 ) \\
  \Phi_{xy} &=&  \rho \, u \, v    \\ 
  \Phi_{qx} &=& - \rho \, \lambda^2\, u   \\
  \Phi_{qy}  &=&   - \rho \, \lambda^2 \, v  \\
  \Phi_{h} &=&   \rho \, \lambda^4  - 3 \, \rho \, \lambda^2 \, (u^2 + v^2)
\end{array} \right. \monend  
and 
\moneq \label{Phi-d2q9}
\Phi(W) = \big(  \Phi_{\varepsilon} ,\,  \Phi_{xx} ,\,  \Phi_{xy} ,\,   \Phi_{qx} ,\,  \Phi_{qy} ,\,  \Phi_{h} \big)^{\rm t} \,. 
\monend
Then the relaxation operates on the six non-conserved moments and it is parameterized by four ``relaxation coefficients''
$ \, s_e ,\, s_\mu ,\, s_q ,\, s_h $.  These coefficients characterize the process of relaxation with 
a family of multiresolution  times. We have
\moneq \label{relaxation-d2q9}
\left\{ \begin{array} {rcl}
\varepsilon^* &=& \varepsilon + s_\varepsilon \, (\Phi_{\varepsilon} - \varepsilon ) \\ 
 p_{xx}^* &=& p_{xx} + s_\mu  \, ( \Phi_{xx}  - p_{xx}  ) \\ 
 p_{xy}^* &=& p_{xy} + s_\mu  \, ( \Phi_{xy}  - p_{xy}  ) \\
 q_x^* &=& {q_x} + s_{q}  \, ( \Phi_{qx}  - q_x  ) \\ 
 q_y^* &=& {q_y} + s_{q}  \, ( \Phi_{qy}  - q_y  ) \\
{h}^* &=& {h} + s_{h}  \, ( \Phi_{h}  - {h}  ) \, .
\end{array} \right. \monend 
We write the previous relations in a compact vector form.
We first introduce a diagonal matrix containing all the relaxation coefficients:
\moneq \label{S-d2q9}
S  = {\rm diag} \, \big(  s_\varepsilon ,\, s_\mu ,\, s_\mu ,\, s_q ,\, s_q ,\, s_h \big) \, .
\monend
For the approximation ``BGK'' of the Boltzmann equation initially introduced  by
Bhatnagar, Gross and Krook  \cite{BGK54}, 
all relaxation coefficients are identical.

\smallskip \noindent 
Then the relations (\ref{relaxation-d2q9}) can be written 
\moneq \label{Y-to-Ystar-d2q9}
Y^* = Y +    S  \, ( \Phi(W) - Y)  \, .
\monend 
At the end of the relaxation process, we have constructed the  vector $ \, m^* \, $ of
``moments after relaxation'': 
\moneqstar 
m^* \equiv \begin{pmatrix} W^* \\ Y^* \end  {pmatrix} =
\begin{pmatrix} W \\  Y + S  \, ( \Phi(W) - Y )   \end  {pmatrix} 
\monendstar 
and the vector of equilibria $ \, \Phi(W) \, $ is evaluted thanks to the relation
(\ref{Phi-d2q9}).

\bigskip

\noindent 
The second step of one iteration of a MRT lattice Boltzmann scheme is the  linear advection process.
Once the  vector $ \, m^* \, $ of moments after relaxation is define, it is transformed into particles populations: 
\moneqstar 
f^{*} = M^{-1} \,  m^{*} \, .
\monendstar 
Then these particles are advected with the velocities $ \, e_j \, $ of the scheme:
\moneq \label{advection-d2q9}
f_j(x,\, t+\Delta t) \,=\, f_j^*(x - e_j \, \Delta t , \, t) 
\monend 
which  corresponds to the method of characteristics
for the advection equation
\moneqstar
{{\partial f_j}\over{\partial t}} + e_j^\alpha \, {{\partial f_j}\over{\partial x^\alpha}} = 0 
\monendstar 
when it is exact!

\bigskip \bigskip    \noindent {\bf \large    3) \quad  {Equivalent partial differential equations}}

\smallskip \noindent
The  ``ABCD'' method for deriving the asymptotic equilavent partial differential equations
for the macroscopic moments 
(Dubois \cite{Du22}) corresponds to a mathematical reformulation
of the discrete multiscale Chapman-Enskog analysis
proposed by Alexander {\it et al.} \cite{ACS93}, 
 McNamara and Alder \cite{McNA93},  
 Qian and Zhou~\cite{QZ00}, as established in Dubois {\it et al.} \cite{DBL23}.

 \noindent 
We first introduce  the  advection operator in the basis of moments  
\moneqstar
\Lambda =  M \, {\rm diag} \Big(  \sum_{1  \leq \alpha \leq d}  e_\alpha \, \partial_\alpha \Big) \, M^{-1} \, . 
\monendstar
For the D2Q9 lattice Boltzmann scheme and the d'Humi\`eres matrix $ \, M \, $ proposed in Eq.~(\ref{M-d2q9}), we have
\moneqstar
\Lambda = 
\left[ \begin{array}{|ccc|cccccc|} \hline 
 0 & \!\! \partial_x & \!\! \partial_y &  0 & 0 & \!\!  0 & \!\! 0 & 0 & 0  \\ 
{{2\lambda^2}\over{3}} \, \partial_x & \!\!  0 & \!\!  0 & {1\over6} \, \partial_x & 
{1\over2} \, \partial_x  &  \!\!  \partial_y  & \!\! 0 & 0  &  0 \\
{{2\lambda^2}\over{3}} \, \partial_y & \!\!  0 & \!\!  0 & {1\over6} \, \partial_y & 
  -{1\over2} \, \partial_y  &   \!\!  \partial_x & \!\!  0  &  0 & 0 \\
  \vspace{-.4 cm} & & & & & & & & \\ \hline  \vspace{-.4 cm} & & & & & & & &  \\
 0 & \!\!  \lambda^2 \, \partial_x  & \!\!  \lambda^2 \, \partial_y  &  0 & 0 & \!\! 0 &  \!\!  \partial_x  &    \partial_y  & 0 \\ 
0 &  \!\!  {{\lambda^2}\over{3}} \, \partial_x  &  \!\! -{{\lambda^2}\over{3}} \, \partial_y  &  0 & 0  & \!\! 0 & \!\!  - {1\over3} \, \partial_x  & 
 {1\over3} \, \partial_y   &  0  \\ 
0 &  \!\!  {{2 \, \lambda^2}\over{3}} \, \partial_y  &  \!  {{2 \,\lambda^2}\over{3}} \, \partial_x  & \!\!   0 &  0 & \!\! 
0 & \!\!   {1\over3} \, \partial_y   &  \!\!   {1\over3} \, \partial_x  & 0 \\ 
0 & \!\!  0 & \!\! 0 &   {{\lambda^2}\over{3}} \, \partial_x  &  - \lambda^2 \, \partial_x  & \!\!  \lambda^2 \, \partial_y  &
\!\! 0 & 0 &  {1\over3} \, \partial_x   \\ 
0  & \!\!  0 & \!\!  0 &    {{\lambda^2}\over{3}} \, \partial_y  &    \lambda^2 \, \partial_y  &  \lambda^2 \, \partial_x   & \!\! 0 &
0  & \!\!    {1\over3} \, \partial_y  \\ 
0 & \!\!  0  & \!\! 0 & 0 &   0  & 0 & \lambda^2 \, \partial_x  & \lambda^2 \, \partial_y  &  0  \\ \hline 
  \end{array}    \right] \, . 
\monendstar 
This matrix is decomposed into four blocks:
\moneq \label{abcd-d2q9}
\Lambda = \begin{pmatrix} A & B \\ C & D \end{pmatrix}
\monend
with four matrices $\, A $, $ \, B $, $ \, C \, $ and $ \, D \, $  of differential operators associated to the decomposition
proposed in Eq.~(\ref{m-WY-d2q9})
of the moments:
\moneqstar
A = \begin{pmatrix}  0 &  \partial_x & \partial_y \\ {{2\lambda^2}\over{3}} \, \partial_x & 0 & 0 \\
  {{2\lambda^2}\over{3}} \, \partial_y & 0 & 0 \end{pmatrix} \,, \quad 
B = \begin{pmatrix} 0 & 0 & 0 & 0 & 0 & 0 \\
  {1\over6} \, \partial_x & {1\over2} \, \partial_x  &  \partial_y  & 0 & 0  &  0 \\ 
  {1\over6} \, \partial_y &   -{1\over2} \, \partial_y  & \partial_x & 0  &  0 & 0 \end{pmatrix} \,, 
\monendstar 

\moneqstar
C = \begin{pmatrix}  0 & \lambda^2 \, \partial_x  &  \lambda^2 \, \partial_y  \\
0 &   {{\lambda^2}\over{3}} \, \partial_x  &  -{{\lambda^2}\over{3}} \, \partial_y  \\ 
0 &   {{2 \, \lambda^2}\over{3}} \, \partial_y  &   {{2 \,\lambda^2}\over{3}} \, \partial_x \\ 
0 & 0 & 0 \\ 0 & 0 & 0 \\ 0 & 0 & 0 \end{pmatrix} \,, \,\, 
D = \begin{pmatrix}  0 & 0 & 0 &  \partial_x  &    \partial_y & 0 \\ 
  0 & 0 & 0 & - {1\over3} \, \partial_x  &  {1\over3} \, \partial_y   &  0 \\ 
  0 & 0 & 0 & {1\over3} \, \partial_y   &   {1\over3} \, \partial_x  & 0 \\ 
  {{\lambda^2}\over{3}} \, \partial_x  &  - \lambda^2 \, \partial_x  &  \lambda^2 \, \partial_y  &
  0 & 0 &  {1\over3} \, \partial_x \\ 
 {{\lambda^2}\over{3}} \, \partial_y  &    \lambda^2 \, \partial_y  &  \lambda^2 \, \partial_x   &  0 &
0  &    {1\over3} \, \partial_y \\ 
 0 & 0 & 0 &  \lambda^2 \, \partial_x  & \lambda^2 \, \partial_y  &  0   
\end{pmatrix} \,. 
\monendstar 
Then, as remarked in \cite{Du22}, 
each  iteration  (\ref{advection-d2q9})  of the lattice Boltzmann scheme
can be expressed as an exact exponential expression: 
\moneq \label{m-exponentielle-m-star}
m  (x, t + \Delta t) =  {\rm exp} ( - \Delta t\, \Lambda ) \,\,  m^*(x ,\, t) \, .
\monend
In practice, we must consider the development of the exponential of an operator:
\moneqstar
\exp( -\Delta t \,  \Lambda) = {\rm I} - \Delta t  \, \Lambda + {{\Delta t^2}\over{2}} \,  \Lambda^2 +   
    {\rm O}(\Delta t^3) \, . 
\monendstar 
Moreover,  we have to be careful with the non commutation of the product of two matrices,
even if all the  partial differential operator commute! For example, we have 
\moneqstar
\Lambda^2 =   \Lambda \, \,   \Lambda   
 = \begin{pmatrix} A  &  B \\  C &   D  \end  {pmatrix} \begin{pmatrix} A  &  B \\  C &   D  \end  {pmatrix} 
 = \begin{pmatrix}   A^2  + B \, C  & A  \, B  + B \,  D \\
 C  \, A + D \, C   & C \, B  + D^2   \end  {pmatrix} \, . 
\monendstar 
With a formal expansion of the evolution into several scales: 
\moneqstar
  \partial_t = \partial_{t_1}  + \Delta t \, \partial_{t_2} + {\rm O}(\Delta t^2)   \,, 
\monendstar 
the Taylor expansion with ABCD  method  revisit the multiscale Chapman-Enskog. 
The  equivalent partial differential equations  of the scheme 
can be written 
\moneq \label{equivalent-edp-ordre-2}
\left\{ \begin{array} {l}
\partial_{t_1} W + \Gamma_1 = 0 \\  \partial_{t_2} W +   \Gamma_2  = 0  \\ 
\Gamma_1 =   A  \, W + B  \, \Phi(W)  \\ 
 Y = \Phi(W) +  \Delta t \,  S^{-1} \, \Psi_1  + {\rm O}(\Delta t^2)  \\ 
 \Psi_1 =  \dd \Phi(W) .  \Gamma_1  - ( C\, W + D \, \Phi(W)) \\ 
 \Sigma  \equiv  S^{-1} - {1\over2} \, {\rm I} \\ 
\Gamma_2 =  B \,  \Sigma \, \Psi_1   \, . 
\end{array} \right. \monend
In the relations (\ref{equivalent-edp-ordre-2}), we have introduced the diagonal
matrix $ \, \Sigma $. 
This matrix  is called the H\'enon matrix \cite{He87}. For the D2Q9 scheme, we have 
\moneq \label{henon} 
\Sigma =  {\rm diag} \, ( \sigma_e,\, \sigma_\mu,\, \sigma_\mu,\, \sigma_q,\, \sigma_q,\,  \sigma_h  ) 
\monend
and in particular
\moneq  \label{sigma_e-sigma_x} 
\sigma_e = {{1}\over{s_e}} - {1\over2} \,,\,\, \sigma_\mu = {{1}\over{s_\mu}} - {1\over2} \, . 
\monend
%

\bigskip

\noindent
As a reult of the previous asymptotic analysis, the  isothermal Navier-Stokes  emerge
at second order when the third order terms relative to the velocity are neglected
(see {\it e.g.} \cite{De03,De14,Du22,GSPK15,JKL05,LL00} and many other references).
The asymptotic model  satisfies the conservation of mass and momentum:
\moneq  \label{NS-isotherme-2d} 
\left\{ \begin{array} {l}
  \partial_t \rho + \partial_x(\rho \, u) + \partial_y (\rho \, v) = 0 \\
  \partial_t (\rho \, u)  + \partial_x (\rho \, u^2 + p) + \partial_y (\rho \, u \, v) =  \partial_x \tau_{xx} + \partial_y \tau_{xy} \\ 
  \partial_t (\rho \, v) + \partial_x (\rho \, u \, v) + \partial_y(\rho \, v^2 + p) = \partial_x \tau_{xy} + \partial_y \tau_{yy} \, . 
\end{array} \right. \monend
The pressure is given by the relation $ \,  p = c_0^2 \, \rho \, $ with
the speed of sound $ \, c_0 \, $ obtained by the relation 
$ \, c_0 = {{\lambda}\over{\sqrt{3}}} $. The viscous tensor in right hand side of (\ref{NS-isotherme-2d})
satisfies 
\moneqstar
\left\{ \begin{array} {l}
\tau_{xx}  =  2 \,\mu \,  \partial_x u   + (\zeta-\mu)\,   ( \partial_x u  +  \partial_y v)  \\
\tau_{xy}  = \mu  \, (\partial_x v  + \partial_y u ) \\ 
\tau_{yy}  =   (\zeta-\mu) ( \partial_x u  +  \partial_y v)  + 2 \, \mu  \, \partial_y v  \, .
\end{array} \right. \monendstar 
The shear  viscosity $ \, \mu \, $ satisfies
$ \, \mu =  {{\lambda}\over{3}} \,  \rho \, \sigma_\mu \, \Delta x   \, $
and the bulk viscosity  $ \,   \zeta \, $ is given by the relation $ \, \zeta = {{\lambda}\over{3}} \,  \rho  \, \sigma_e \, \Delta x  $.

\noindent 
The preceeding algebraic  relations are valid only for two space dimensions.
For $ \, D \, $ space dimensions, we have 
\moneqstar
\tau_{\alpha  \beta}  =  \mu \,  \big( \partial_\alpha u_\beta + \partial_\beta u_\alpha \big)
  + \Big( \zeta- {{2}\over{D}} \, \mu \Big) \,   ( {\rm div} \, {\bf u} ) \, \delta _{\alpha  \beta}
\monendstar
and the trace of the viscous stress is given by $ \, \tau_{\alpha  \alpha} = \zeta \, D \,( {\rm div} \, {\bf u} ) \, $
with $ \, \zeta \, $ the bulk viscosity. 
Moreover,  with the usual definition of viscous stress tensor employed in kinetic theory, 
where this tensor is interpreted as the macroscopic flux of momentum, a minus sign arises.
It differs from the convention
used in the Fluid Mechanics theory, where the tensor represents the stress exerted on the fluid by
neighboring fluid particles and the minus sign is absent.

\bigskip

\noindent 
We observe that the equilibrium value of the 
viscous  moments $ \, \varepsilon,\, p_{xx} ,\, p_{xy} \, $  fix perfect fluid terms. 
For this reason, the vector $ \, Y_e\, $ of viscous moments is  defined by 
\moneq \label{moments-Ye-d2q9}
 Y_e = \begin{pmatrix} \varepsilon  \\  p_{xx} \\  p_{xy} \end {pmatrix}  .
\monend
In a similar way, the equilibrium value of the energy transfer moments $ \, q_x,\, q_y \, $ allow adjustment of second-order terms.
In this contribution,   the  energy transfer moments  $ \, Y_v \, $ are given according to 
\moneq \label{moments-Yv-d2q9}
 Y_v = \begin{pmatrix} q_x  \\  q_y \\ h \end {pmatrix}  . 
\monend
The expression of the equilibrium of the ``ghost moment'' $ \, h \, $ has been studied by Lallemand and Luo \cite{LL00},
Dellar \cite {De03, De14}, Geier \cite{GSPK15} among others.
We observe here that 
this last moment has no impact on the second order equations  (\ref{NS-isotherme-2d}).

\bigskip \bigskip    \noindent {\bf \large    4) \quad  {Projected multiple relaxation time D2Q9 scheme}}

\smallskip \noindent
Following the remark done at the end of the previous section, we replace the family $ \, Y \, $
of non-conserved moments by two families $ \, Y_e \, $ and $ \, Y_v \, $ with 
$ \, Y = ( Y_e ,\, Y_v )^{\rm t} $. 
The important point concerns the  advection operator  $ \, \Lambda \, $  in the basis of moments.
For  isothermal  D2Q9  studied in the previous section, 
the  operator  $ \, \Lambda \, $ contains a certain number of zero blocks, as observed therehein: 
\moneqstar
\Lambda = 
\left[ \begin{array}{|ccc|ccc|ccc|} \hline 
 0 & \!\! \partial_x & \!\! \partial_y &  0 & 0 & \!\!  0 & \!\! 0 & 0 & 0  \\ 
{{2\lambda^2}\over{3}} \, \partial_x & \!\!  0 & \!\!  0 & {1\over6} \, \partial_x & 
{1\over2} \, \partial_x  &  \!\!  \partial_y  & \!\! 0 & 0  &  0 \\
{{2\lambda^2}\over{3}} \, \partial_y & \!\!  0 & \!\!  0 & {1\over6} \, \partial_y & 
  -{1\over2} \, \partial_y  &   \!\!  \partial_x & \!\!  0  &  0 & 0 \\
  \vspace{-.4 cm} & & & & & & & & \\ \hline  \vspace{-.4 cm} & & & & & & & &  \\
 0 & \!\!  \lambda^2 \, \partial_x  & \!\!  \lambda^2 \, \partial_y  &  0 & 0 & \!\! 0 &  \!\!  \partial_x  &    \partial_y  & 0 \\ 
0 &  \!\!  {{\lambda^2}\over{3}} \, \partial_x  &  \!\! -{{\lambda^2}\over{3}} \, \partial_y  &  0 & 0  & \!\! 0 & \!\!  - {1\over3} \, \partial_x  & 
 {1\over3} \, \partial_y   &  0  \\ 
0 &  \!\!  {{2 \, \lambda^2}\over{3}} \, \partial_y  &  \!  {{2 \,\lambda^2}\over{3}} \, \partial_x  & \!\!   0 &  0 & \!\! 
0 & \!\!   {1\over3} \, \partial_y   &  \!\!   {1\over3} \, \partial_x  & 0 \\
  \vspace{-.4 cm} & & & & & & & & \\ \hline  \vspace{-.4 cm} & & & & & & & &  \\
0 & \!\!  0 & \!\! 0 &   {{\lambda^2}\over{3}} \, \partial_x  &  - \lambda^2 \, \partial_x  & \!\!  \lambda^2 \, \partial_y  &
\!\! 0 & 0 &  {1\over3} \, \partial_x   \\ 
0  & \!\!  0 & \!\!  0 &    {{\lambda^2}\over{3}} \, \partial_y  &    \lambda^2 \, \partial_y  &  \lambda^2 \, \partial_x   & \!\! 0 &
0  & \!\!   {1\over3} \, \partial_y  \\ 
0 & \!\!  0  & \!\! 0 & 0 &   0  & 0 & \lambda^2 \, \partial_x  & \lambda^2 \, \partial_y  &  0  \\ \hline 
  \end{array}    \right] \, .
\monendstar
In other terms, we have a 3 by 3 structure composed by six non-zero blocks of 3 by 3 matrices:
\moneq \label{Lambda-d2q9-blocs}
\Lambda =  \begin{pmatrix} A   &  B_e &   0 \\
  C_e  & 0 &  D_{ev}  \\ 0 &  D_{ve}  &  D_{vv}  \end  {pmatrix} 
\monend
with 
\moneqstar
A = \begin{pmatrix}  0 &  \partial_x & \partial_y \\ {{2\lambda^2}\over{3}} \, \partial_x & 0 & 0 \\
  {{2\lambda^2}\over{3}} \, \partial_y & 0 & 0 \end{pmatrix} , \,\, 
B_e = \begin{pmatrix} 0 & 0 & 0  \\
  {1\over6} \, \partial_x & {1\over2} \, \partial_x  &  \partial_y   \\ 
  {1\over6} \, \partial_y &   -{1\over2} \, \partial_y  & \partial_x  \end{pmatrix} , \,\, 
C_e = \begin{pmatrix}  0 & \lambda^2 \, \partial_x  &  \lambda^2 \, \partial_y  \\
0 &   {{\lambda^2}\over{3}} \, \partial_x  &  -{{\lambda^2}\over{3}} \, \partial_y  \\ 
0 &   {{2 \, \lambda^2}\over{3}} \, \partial_y  &   {{2 \,\lambda^2}\over{3}} \, \partial_x \end{pmatrix} , \,\, 
\monendstar 
\moneqstar
D_{ev} =  \begin{pmatrix} \partial_x  &    \partial_y & 0 \\ - {1\over3} \, \partial_x  &  {1\over3} \, \partial_y   &  0 \\ 
   {1\over3} \, \partial_y   &   {1\over3} \, \partial_x  & 0  \end{pmatrix} , \,\,  
D_{ve} =  \begin{pmatrix}   {{\lambda^2}\over{3}} \, \partial_x  &  - \lambda^2 \, \partial_x  &  \lambda^2 \, \partial_y  \\ 
 {{\lambda^2}\over{3}} \, \partial_y  &    \lambda^2 \, \partial_y  &  \lambda^2 \, \partial_x     \\ 
 0 & 0 & 0   \end{pmatrix} , \,\,
D_{vv} =  \begin{pmatrix}    0 & 0 &  {1\over3} \, \partial_x \\ 0  &  0  &  {1\over3} \, \partial_y \\
  \lambda^2 \, \partial_x  & \lambda^2 \, \partial_y  &  0  \end{pmatrix} . 
\monendstar 
In consequence, it is natural to propose a 
new structure for the moments  with three components:
\moneq \label{moments-d2q9-3-blocs}
m =  \begin{pmatrix}  W \\  Y_e \\  Y_v \end  {pmatrix}  .
\monend
First  the conserved moments 
\moneq \label{moments-W-d2q9}
W = \begin{pmatrix} \rho \\ \rho \, u \\ \rho \, v \end  {pmatrix}  ,
\monend
then the non conserved moments $ \, Y \, $ decomposed into two sub-families:
\moneq \label{moments-Y-d2q9}
 Y = \begin{pmatrix} Y_e \\  Y_v  \end {pmatrix}  
 \monend
with the Eulerian moments $ \, Y_e \, $ introduced in (\ref{moments-Ye-d2q9})
and the viscous moments $ \, Y_v \, $ defined at the relation  (\ref{moments-Yv-d2q9}). 
With this new sub-structure, the  non conserved moments at equilibrium can be written
\moneqstar 
\Phi (W) =  \begin{pmatrix}  \Phi_e  \\  \Phi_v   \end {pmatrix} .
\monendstar
The eulerian moments at equilibrium $ \, \Phi_e \, $ are obtained with the usual relations (\ref{formules-Phi-d2q9}): 
%
\moneq \label{moments-Phie-d2q9}
\Phi_e   = \begin{pmatrix}
\Phi_{\varepsilon}   \\  \Phi_{xx}  \\  \Phi_{xy} \end{pmatrix} =
\begin{pmatrix}
  -  2 \, \lambda^2 \, \rho \, + \, 3 \, \rho \, ( u^2 + v^2 )  \\
  \rho \, ( u^2 - v^2 )  \\ 
  \rho \, u \, v  \end{pmatrix}
\monend
and we take for the viscous moments at equilibrium  $ \, \Phi_v \, $ 
(see again (\ref{formules-Phi-d2q9})): 
\moneq \label{moments-Phiv-d2q9}
\Phi_v  = \begin{pmatrix}
\Phi_{qx}  \\ \Phi_{qy} \\ \Phi_{h}  \end{pmatrix} = 
\begin{pmatrix}
 - \rho \, \lambda^2\, u  \\ 
 - \rho \, \lambda^2 \, v  \\
 \rho \, \lambda^4  - 3 \, \rho \, \lambda^2 \, (u^2 + v^2)   \end{pmatrix}  . 
\monend
The defect of equilibrium $ \, \Psi \, $ is also  decomped into two components:
\moneq \label{psi1-d2q9-v2}
\Psi_1 =  \begin{pmatrix}  \Psi_e \\ \Psi_v   \end {pmatrix} .
\monend 
Recall that the H\'enon matrix   $ \, \Sigma \, $ is defined according to 
\moneq \label{Sigma-d2q9}
\Sigma  =   S^{-1} - {1\over2} \, {\rm I} . 
\monend
This diagonal matrix is now considered as decomposed into two blocks:
\moneq \label{Sigma-d2q9-v2}
\Sigma  = {\rm diag} \, \big( \,  \Sigma_e  \,,\, \Sigma_v \,\big)
\monend
and the first block $ \,  \Sigma_e \, $ is a diagonal 3 by 3 matrix:
\moneq \label{Sigmae-d2q9}
\Sigma_e =  {\rm diag} \, ( \sigma_e,\, \sigma_\mu,\, \sigma_\mu  )  .
\monend
with $ \, \sigma_e \, $ and $ \, \sigma_\mu \, $ specified in relation (\ref{sigma_e-sigma_x}).

\bigskip \bigskip    \noindent {\bf \large    5) \quad  {Asymptotic analysis of the MRT scheme with projection}} 

\smallskip \noindent
We first revisit the classic MRT scheme when the struture emerging in relations
(\ref{Lambda-d2q9-blocs})-(\ref{Sigmae-d2q9}) 
is taken into account. 
With this choice of a substruture, and in particular the advection matrix in the basis
of moments given according to the block decomposition (\ref{Lambda-d2q9-blocs}), 
the equivalent partial differential equations of the second order of the MRT scheme
described by  the relations~(\ref{equivalent-edp-ordre-2}) take the form 
\moneq \label{equivalent-edp-ordre-2-v2}
\left\{ \begin{array} {l}
\partial_{t_1} W + \Gamma_1 = 0 \\  
\Gamma_1 =   A  \, W + B_e  \, \Phi_e(W)  \\ 
 Y_e = \Phi_e(W) +  \Delta t \,  S_e^{-1} \, \Psi_e  + {\rm O}(\Delta t^2)  \\ 
 \Psi_e =  \dd \Phi_e(W) .  \Gamma_1  - ( C_e  \, W + D_{ev} \, \Phi_v(W)) \\ 
 \Sigma_e  = {\rm diag} (\sigma_e,\, \sigma_\mu ,\, \sigma_\mu ) \\ 
 \Gamma_2 =  B_e \,  \Sigma_e \, \Psi_e   \\
 \partial_{t_2} W +   \Gamma_2  = 0  \, .  
\end{array} \right. \monend
First observe that the relations (\ref{equivalent-edp-ordre-2-v2}) are
completely equivalent to the initial system (\ref{equivalent-edp-ordre-2}).
This property can be demonstrated as follows.
We have from (\ref{equivalent-edp-ordre-2}) the relation 
$ \,  \partial_{t_1} W +  \Gamma_1  = 0 $.
Then  the following calculus
\moneqstar
\Gamma_1 =  A \, W + B  \, \Phi(W) =      A  \, W +  B_e  \, \Phi_e(W)
\monendstar
establishes  the first order relations in (\ref{equivalent-edp-ordre-2-v2}).
For the nonconserved moments, the relation 
$ \, Y = \Phi(W) +  \Delta t \,  S^{-1} \, \Psi_1  + {\rm O}(\Delta t^2) \, $
is splitted into two components according to (\ref{moments-Y-d2q9}).
For the first component, we have
$ \, Y_e = \Phi_e(W) +  \Delta t \,  S_e^{-1} \,  \Psi_e  + {\rm O}(\Delta t^2)  $.
Now, the two sub-vectors decomposition (\ref{psi1-d2q9-v2}) introduces a first component
$ \,  \Psi_e $. from the structure (\ref{Lambda-d2q9-blocs}),
we deduce 
\moneqstar
\Psi_e  =  \dd \Phi_e (W) .  \Gamma_1 -  ( C_e \, W +  D_{ev}  \, \Phi_v (W)) .
\monendstar
Then, due to the substructuring of the H\'enon matrix, we have (\ref{Sigma-d2q9-v2})
\moneqstar
\Gamma_2 =  B \,  \Sigma \, \Psi_1  = B_e \,  \Sigma_e \, \Psi_e
\monendstar
and the set of relations (\ref{equivalent-edp-ordre-2-v2}) is established. \hfill $\square$

\bigskip

We observe now that, as previously observed in \cite{LC06,Ma15,MPH17} 
in an other context, the  equilibria   (\ref{moments-Phiv-d2q9}) are related to
the equilibria (\ref{moments-Phie-d2q9}) according to the relation
\moneq \label{lien-Phiv-Phie}
\Phi_v   =  K \, W + L \,  \Phi_e
\monend
with 
\moneq \label{K-L}
K = \begin{pmatrix} 0 & - \lambda^2 & 0 \\ 0 & 0 & - \lambda^2 \\
  - \lambda^4 & 0 & 0 \end{pmatrix} \,,\quad  
L = \begin{pmatrix} 0 & 0 &0 \\ 0 & 0 & 0 \\   - \lambda^2 & 0 & 0 \end{pmatrix}  \, .
\monend
The relation (\ref{lien-Phiv-Phie}) has been inspired by the recurrence
relations between Hermite polynomials, developed in Malaspinas \cite{Ma15}
and Mattila {\it et al.} \cite{MPH17}. 
Observe that we have simply
\moneqstar  \left\{ \begin{array} {l}
- \rho \, \lambda^2\, u   =  - \lambda^2 \, ( \rho \, u) \\
 - \rho \, \lambda^2\, v  =  - \lambda^2 \, ( \rho \, v) \\
 \rho \, \lambda^4  - 3 \, \rho \, \lambda^2 \, (u^2 + v^2)    = - \lambda^4 \, \rho 
 - \lambda^2 \, \big( -  2 \, \lambda^2 \, \rho \, + \, 3 \, \rho \, ( u^2 + v^2 ) \big)
\end{array} \right. \monendstar
and the relation (\ref{lien-Phiv-Phie}) is a simple consequence of  (\ref{moments-Phie-d2q9})
and  (\ref{moments-Phiv-d2q9}). \hfill $\square$

\smallskip
In conclusion of this important remark, we have obtained with (\ref{lien-Phiv-Phie}) a simple 
expression of the viscous moments at equilibrium $ \,  \Phi_v \,  $
with  a linear expression of the conserved variables~$ \, W \, $
and the eulerian moments $ \, \Phi_e  $.

\bigskip

From the previous remark, we define a projection operator $ \, P \, $ 
in the space of moments by the relation
\moneq \label{projecteur-d2q9}
P \, \begin{pmatrix}  W \\  Y_e \\  Y_v \end  {pmatrix} =
\begin{pmatrix}  W \\  Y_e \\ K \, W + L \, Y_e \end  {pmatrix} 
\monend
with the help of the two matrices $ \, K \, $ and $ \, L \, $ introduced at the relation (\ref{K-L}).
The projected vector $ \, P m \, $ has three compoents:
\moneq \label{Pm-d2q9}
\left\{ \begin{array} {l}
  (Pm)_W = W \\
  (Pm)_e = Y_e \\
  (Pm)_v =  K \, W + L \,  Y_e  \, . 
\end{array} \right. \monend

\bigskip

With this projector operator, we  define 
a multiple relaxation time lattice Boltzmann  scheme with projection      
by the following algorithm. For a set of moments $ \, m \, $ given by the
relation~(\ref{moments-d2q9-3-blocs}) for the D2Q9 lattice Boltzmann scheme, we have 

\smallskip {\it (i)} \qquad
projection of the moments $\, m \longrightarrow Pm $.

Then the moments at equilibrium $ \,  (P  m)^{\rm eq} \, $ can be decomposed into three vector components:
\moneq \label{Pm-equilibre} 
(P  m)^{\rm eq} =   \begin{pmatrix} W \\ \Phi_e \\  K \, W + L \,  \Phi_e   \end{pmatrix}
=   \begin{pmatrix} W \\ \Phi_e \\  \Phi_v   \end{pmatrix} = m^{\rm eq}
\monend

\smallskip {\it (ii)} \quad  relaxation  $\, Pm \longrightarrow  (P  m)^* $. 

We have simply 
\moneq \label{Pm-star} 
(P  m)^* =  \begin{pmatrix} W \\ Y_e^* \\  K \, W + L \,  Y_e^* \end{pmatrix}
\monend
with 
$ \, Y_e^* = ({\rm I} - S_e) \, Y_e + S_e \,  \Phi_e $.
We observe that we have now 
\moneq \label{Yv-star-d2q9}
Y_v^* =  K \, W + L \,  Y_e^*
\monend
instead of $ \, Y_v^* =  ({\rm I} - S_v) \, Y_v + S_v \,  \Phi_v  \, $ for the initial lattice Boltzmann scheme. 

\smallskip {\it (iii)} \quad  propagation

From the moments after relaxation, we introduce the particle representation
\moneq \label{definition-fstar}
f^* = M^{-1}  (P  m)^* .
\monend 
This distribution is exactly advected during one time step:
\moneq \label{propagation}
f_j(x,\, t+\Delta t) \,=\, f_j^*(x - e_j \, \Delta t , \, t) \, .
\monend 
The moments $\, m \, $  at the new time step are a linear transform of the particles: 
$ \, m = M \, f $. Then the algorith can be iterated again.

\bigskip

The  MRT lattice Boltzmann scheme with projection   has the same asymptotic properties at second order than the initial
multiresolution relaxation times  lattice Boltzmann scheme.
An important result of our contribution is  the following 

\bigskip
{\bf  Proposition 1 : a theoretical result  for the MRT scheme with projection}

For the MRT scheme with projection defined at the relations (\ref{projecteur-d2q9})
to   (\ref{propagation}), we have  at second order the following set of partial differential equations
\moneq \label{equivalent-edp-ordre-2-schema-avec -projection}
\left\{ \begin{array} {l}
 \partial_t W + \Gamma_1 + \Delta t \, \Gamma_2 = {\rm O}(\Delta t^2)  \\ 
  \Gamma_1 = A \, W +  B_e  \, \Phi_e   \\
  \Psi_e  =  \dd \Phi_e (W) .  \Gamma_1 - ( C_e  \, W +  D_{ev}  \, \Phi_v (W))  \\
  \Gamma_2 =   B_e  \, \Sigma_e \Psi_e \, . 
\end{array} \right. \monend
At second order of accuracy, the  MRT scheme with projection represents the same physics than the initial
MRT scheme.

\smallskip \noindent 
The proof of this result is developed in Annex 1.

We observe that the resulting model (\ref{equivalent-edp-ordre-2-schema-avec -projection})
is identical to the result (\ref{equivalent-edp-ordre-2-v2}) for 
initial multiresolution relaxation times  lattice Boltzmann scheme.
The projection step does not change the asymptotic physical model at second order!

\bigskip
We observe finally that the preceding proposition is general in scope.
Nevertheless,  we consider in this contribution 
 the MRT scheme with projection only for the D2Q9 lattice Boltzmann scheme.
 If the advection  matrix operator in the basis of moments $ \, \Lambda \, $
admits a structure of the type  (\ref{Lambda-d2q9-blocs}), the equivalent partiel differential
equations satisfy the Proposition~1 and in particular the relations
(\ref{equivalent-edp-ordre-2-schema-avec -projection}).
The projected MRT scheme is derived following a general algorithm. The hypothesis is essentially
that the matrix of advection in the basis of moments admits a block structure
of the type (\ref{abcd-d2q9}) and that the moments at equilibrium verify an
identity of the type (\ref{lien-Phiv-Phie})

\newpage 
\bigskip \bigskip    \noindent {\bf \large    6) \quad  {Numerical experiments for a linear model}} 

\smallskip \noindent
We have implemented the algorithm  (\ref{projecteur-d2q9})-(\ref{propagation}) for the D2Q9 scheme.
Our first results  concern a linearized version of the D2Q9 scheme 
around a constant state with velocity 
\moneqstar
(u_0 ,\, v_0)  = (0.2 ,\, 0 ) 
\monendstar
and sound velocity $ \smash{ c_0 = {{1}\over{\sqrt{3}}} } $.
We have in that case a linearization of the scheme  (\ref{moments-Phie-d2q9})(\ref{moments-Phiv-d2q9})
\moneqstar
W = \begin{pmatrix} \rho \\ \rho \, (u_0 + u) \\ \rho \, v \end{pmatrix} ,\,\, 
\Phi_e (W) = \begin{pmatrix} 
  -  2 \, \lambda^2 \, \rho \, + \, 6 \, \rho \, u_0 \, u \\ 2 \,   \rho \, u_0 \, u  \\ 0   \end{pmatrix} ,\,\, 
\Phi_v (W) = \begin{pmatrix}
 - \rho \, \lambda^2\, (u_0 + u)   \\  - \rho \, \lambda^2 \, v  \\  \rho \, \lambda^4  - 6 \, \rho \, \lambda^2 \, u_0 \, u  \end{pmatrix} .
\monendstar
We can verify very simply that the  matrices $ \, K \, $ and $ \, L \, $ introduced in (\ref{K-L})
satisfy the relation~(\ref{lien-Phiv-Phie}): we have $ \, \Phi_v   =  K \, W + L \,  \Phi_e $. 

\begin{figure}    [H]  \centering

\centerline{{\includegraphics[width=.48\textwidth]{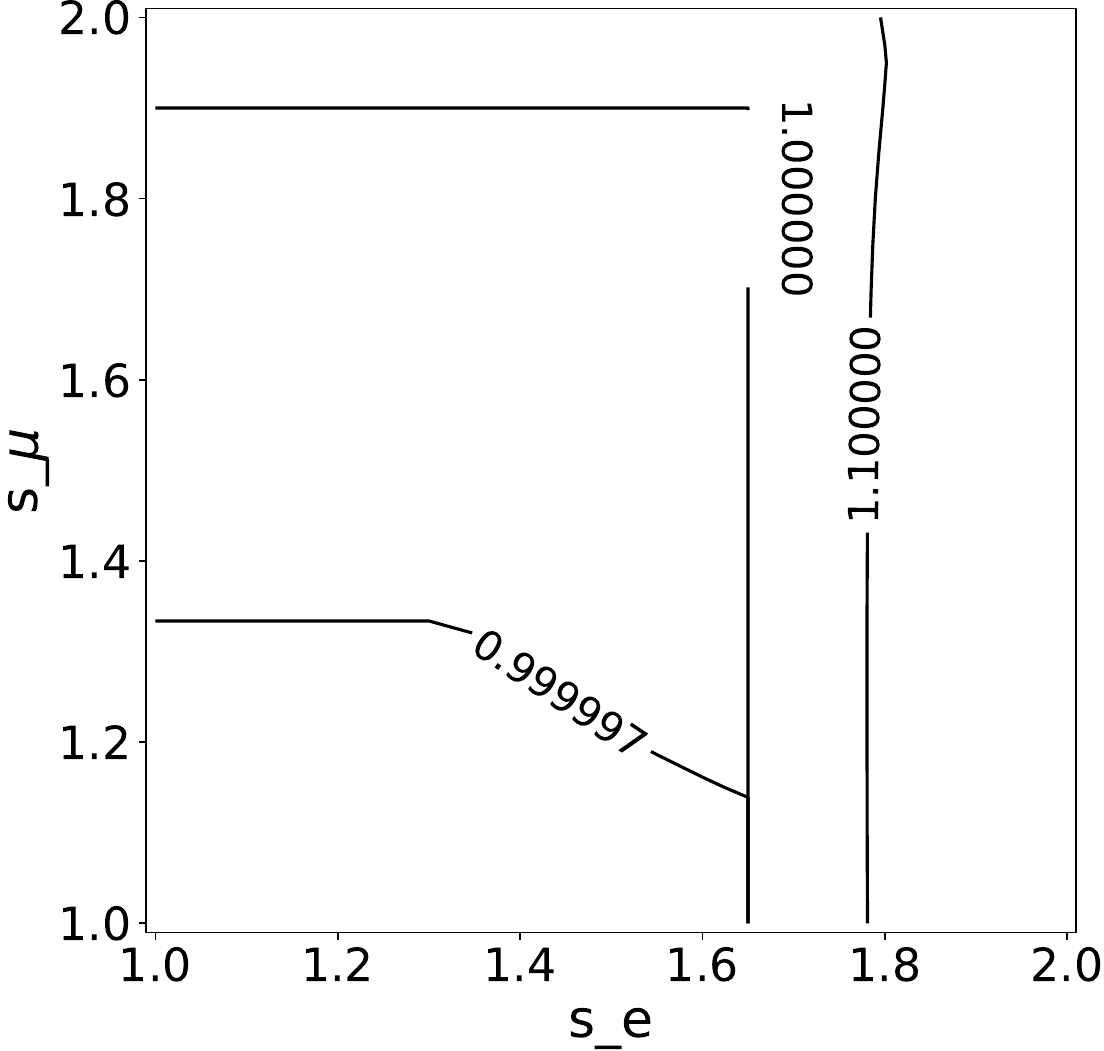}}
  \quad  {\includegraphics[width=.48\textwidth]{ 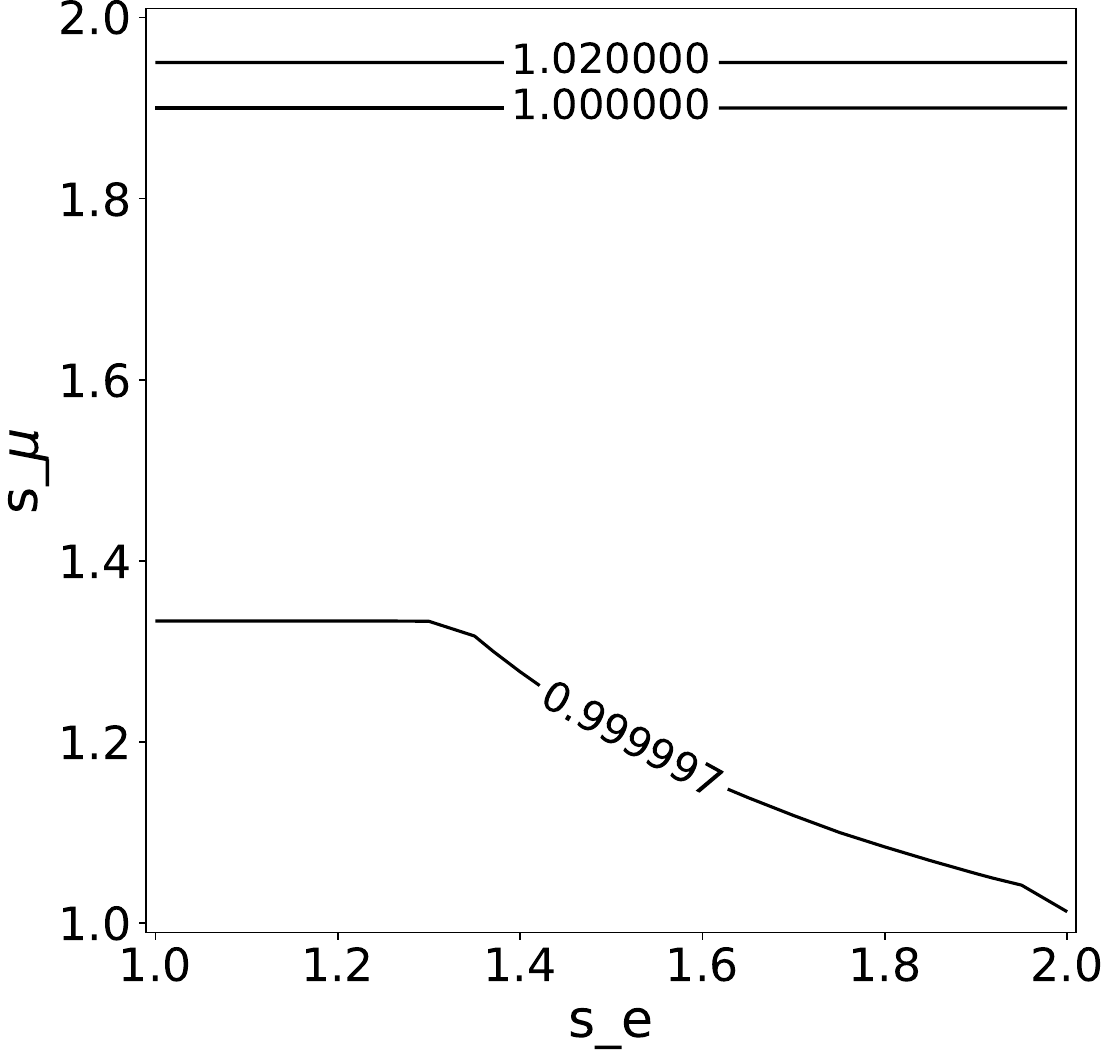}}}
\caption{Comparison of linear stability zones for advection speed $ u_0 = 0.35 \, c_0  $
  and $ \, v_0 = 0 $. 
Traditional D2Q9 scheme \cite{LL00} on the left and D2Q9 MRT with projection on the right.
The stability zone is extended for  higher  values of the relaxation  parameter $\, s_e $.}
\label{fig-01} \end{figure}

\smallskip \noindent 
For each set of parameters ($s_\mu$, $ \, s_e$), we determine the maximum values of the
associated aigenvalues for all the wave vectors, following the method
proposed in Lallemand and Luo~\cite{LL00}. If the maximum of moduli of these eigenvalues is greater than one,
the scheme is unstable. The iso-maxima of the eigenvalues are represented in Figure \ref{fig-01}. 
They show a significant increase in the stability zone
for the relaxation parameter $\, s_e $. 
With the initial MRT scheme, stability is limited to the range
$ \, 0 \leq s_e \leq 1.75 $. The projected MRT lattice Boltzmann scheme
is stable for $ \, 0 \leq s_e \leq 2 \, $ and $ \, 0 \leq s_\mu \leq 1.9 $.

\newpage 
\bigskip \bigskip    \noindent {\bf \large    7) \quad  {Numerical tests for fluid applications}} 

\smallskip \noindent
We have first considered two classical test cases: the Minion-Brown test case \cite {MB97}
studying the performance of under-resolved two-dimensional incompressible flow simulations, and 
the lid-driven  cavity proposed by  Ghia {\it et al.}  \cite {GGS82}

\bigskip 


\smallskip \smallskip
The Minion-Brown  test case describes a  Kelvin-Helmholtz instability.
At the initial time, the density is constant and the velocity is given in the square $ \, [0,\, L]^2 \,$
with $ \, L = 1 \, $ by the relations
\moneq \label{minion-initial}
u = \left\{ \begin{array} {l} \tanh \big[ \kappa \, \big(y - {1\over4} \big) \big] \,\, {\rm {for}} \, \, y \leq {1\over2} \\ 
\tanh  \big[  \kappa \, \big({3\over4} - y \big) \big] \,\, {\rm {for}} \, \, y \geq {1\over2}  \end{array} \right. ,  \,\, 
v = \delta \, \sin \Big[ 2 \pi \big( x + {1\over4} \big) \Big] 
\monend
with
\moneqstar 
\kappa =  80 \,,\,\, 
\delta = 0.05 \, . 
\monendstar 
This test case has been simulated with the help of lattice Boltzmann schemes
by Mari\'e {\it et al.} \cite{MRS09}, Dellar \cite{De14} and Mattila  {\it et al.} \cite{MPH17} among others.

\bigskip

We first explain how we have chosen our numerical parameters. 
As proposed in \cite {MB97}, a relative coarse mesh is used and all our compuatations have been done with
\moneqstar
N = 128
\monendstar 
mesh points in each direction. Then $ \,  \Delta x = {{1}\over{N}} = 0.0078125 $.
We have chosen the same Mach number $\, M_0 \equiv  {{U_0}\over{c_0}} =  0.04 \, $ 
as in reference \cite{MRS09}. 
With the classical value $ \, c_0 = {{1}\over{\sqrt{3}}}\, $ for the sound velocity,
we have a reference velocity $ \, U_0 = 0.0231 $.
Then with the definition $ \, Re \equiv {{U_0 \, L}\over{\nu}} \, $ of the Reynolds number $ \, Re $,
we have $ \, \nu = 2.31 \, 10^{-6} \, $ when $ \, Re = 10^4  $. 
From the classical relation
\moneq \label{nu-sigma-dx}
\nu = \sigma_\mu \, {{\Delta x}\over{3}} 
\monend
(see {\it e. g.} \cite{LL00}), we deduce $ \, \sigma_\mu =  8.8704 \, 10^{-4} \, $ 
and 
\moneq \label{minion-sx}
s_\mu = {{1}\over{0.5 + \sigma_\mu}} = 1.996458123572134  \, .
\monend
The physical duration is fixed to $ \, T =  {{L}\over{U_0}} \approx 43.29 $.
It corresponds to the  final time chosen by Dellar \cite{De14}. 
When $ \, \Delta x = \Delta t $, this value can be approximatively translated  into 
\moneq \label{minion-NT}
N_T =  5541
\monend
iterations of the lattice Boltzmann scheme.

\bigskip

\begin{figure}    [H]  \centering
\centerline {\includegraphics[height=.49 \textwidth]{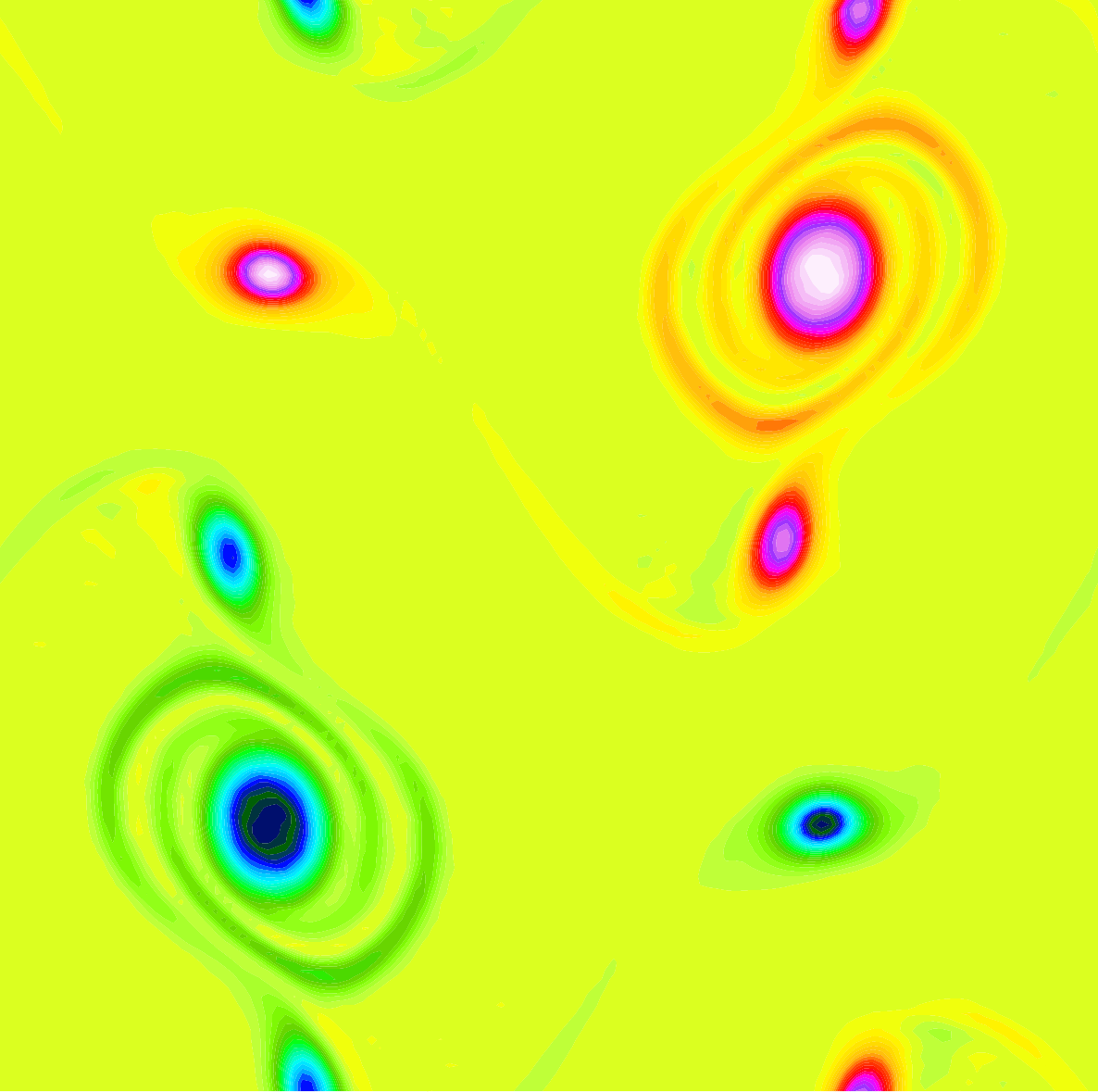}}
\caption{Minion-Brown test case \cite {MB97} for  Reynolds number $ \, Re = 10^4$,
  128 grid points and  $ \, N_T = 5541 \, $ discrete time iterations. 
  BGK results: $\, s_e = s_q = s_h = s_\mu $ given by the relation~(\ref{minion-sx}).
  Vorticity field; the results are not satisfying.}
\label{fig-minion-bgk} \end{figure}

Our first simulation concerns the BGK \cite{BGK54} version of the lattice Boltzmann scheme.
In that case, all the viscosities are taken identical:
\moneq \label{bgk}
s_e = s_q = s_h = s_\mu \, . 
\monend
Curiously, our scheme is not diverging with such parameters. But the results (see Figure \ref {fig-minion-bgk})
have nothing to do with what is physically expected! 

\begin{figure}    [H]  \centering
 {\includegraphics[height=.49 \textwidth]{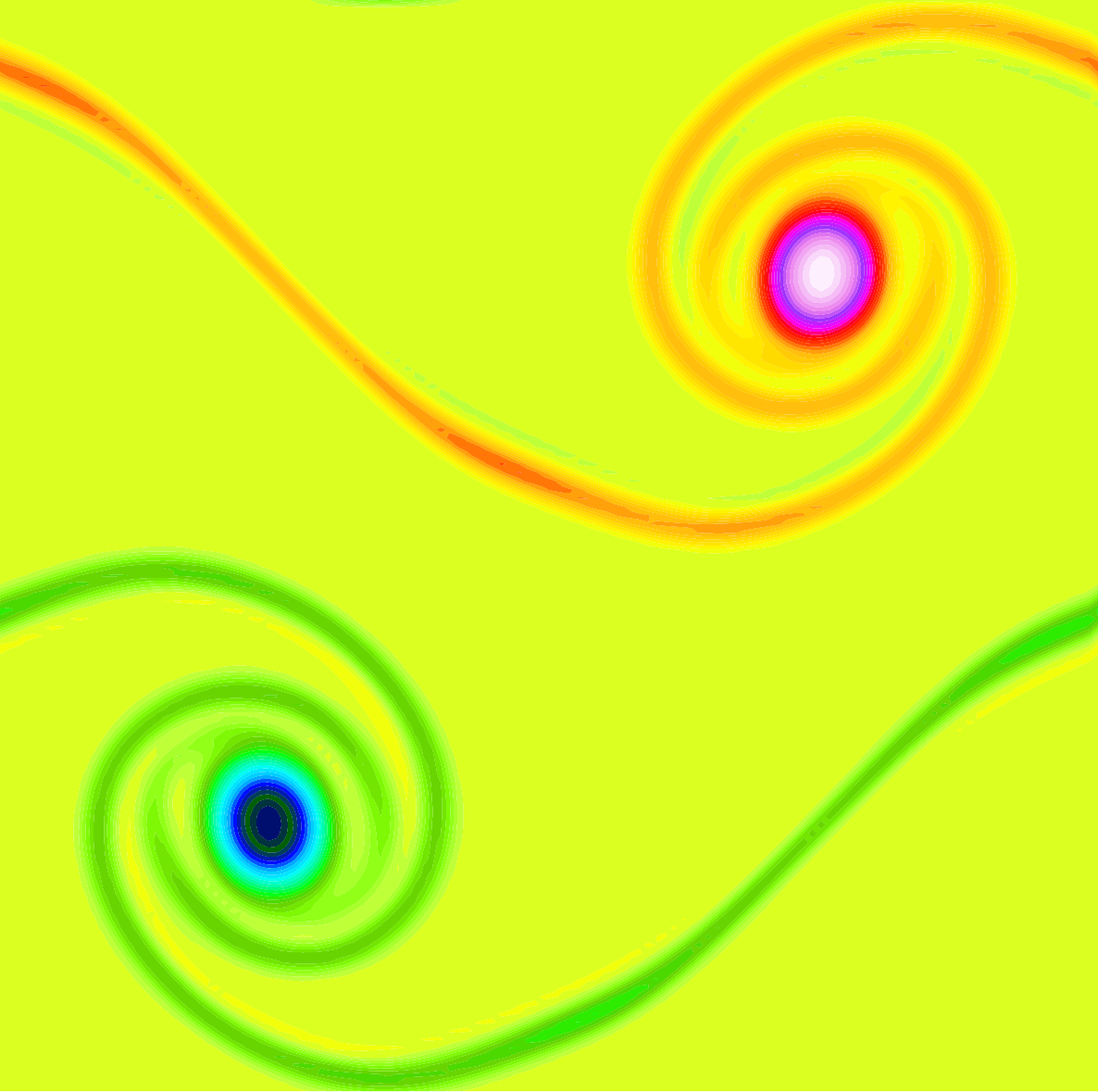}}$\,$ 
 {\includegraphics[height=.49 \textwidth]{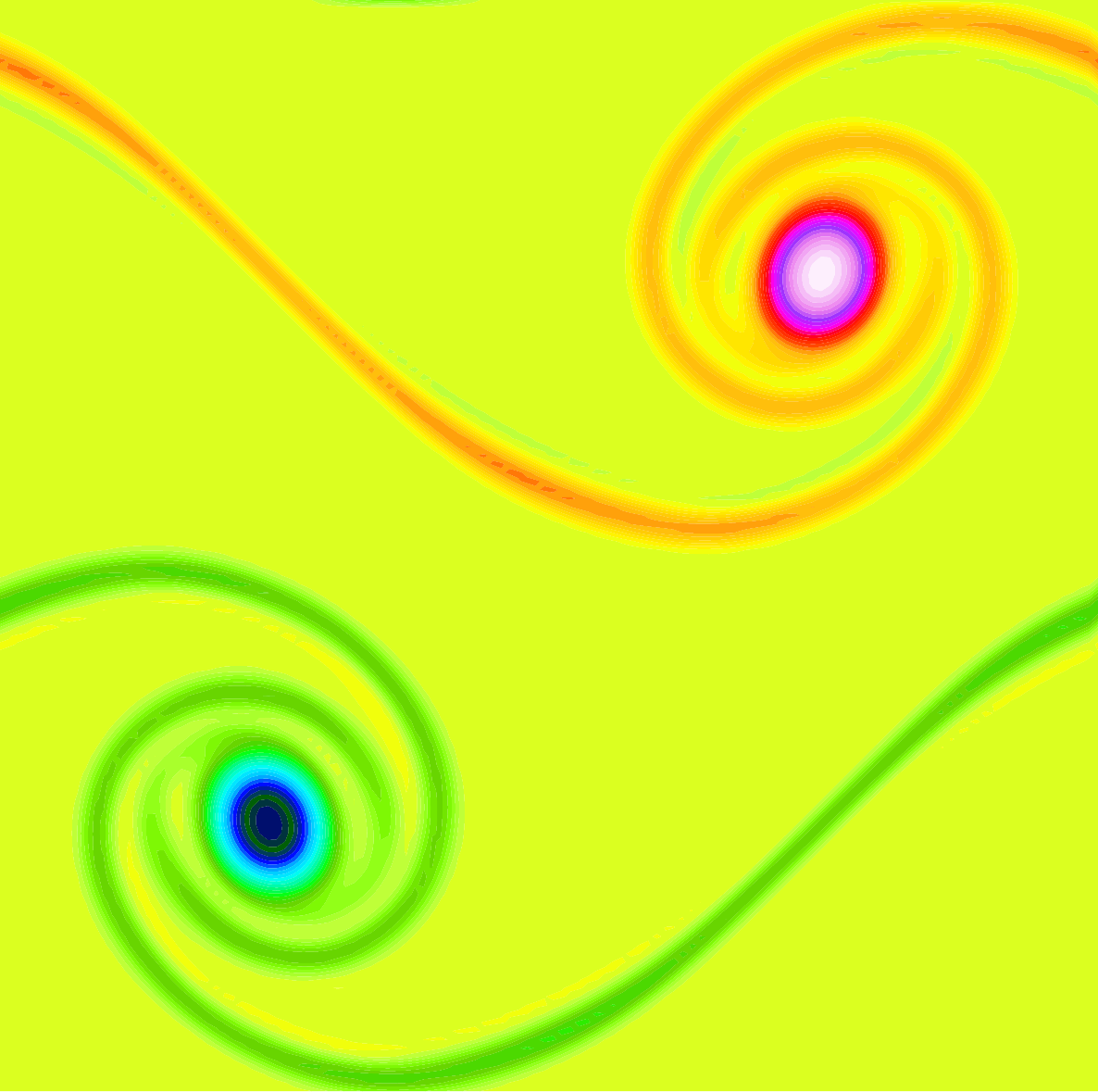}}
\caption{Minion-Brown test case \cite {MB97} for  Reynolds number $ \, Re = 10^4$,
  128 grid points and  $ \, N_T \, $ discrete time iterations (\ref{minion-NT}).
  MRT results on the left with  $\, s_e = 1.715 _, $ and results for the MRT scheme with
  projection on the right with the same parameter $ \, s_e $. 
These vorticity fields are qualitatively correct.}
\label{fig-minion-mrt-projete} \end{figure}

\newpage 
\bigskip

Then we have taken a MRT scheme with $ \, s_q = s_h = 1 \, $ to mimic the effects of the projection scheme.
With the value $ \, s_e = 1.72 $, this MRT scheme is giving overflow values after $ \, N_T \, $ iterations
with the previous parameters.
With $ \, s_e = 1.715 $, the simulation is giving acceptable results.
For this set of parameters, the simulation is close the stability limit for the classic MRT sceme.
The results are presented on the left part
of the Figure \ref{fig-minion-mrt-projete}.
With the MRT lattice Boltzmann scheme with projection, with the same numerical parameters $\, s_e \, $ and $ \, s_\mu $,
the computation does not encounter any difficulty.
The results are close to the MRT ones and are presented on the right of Figure \ref{fig-minion-mrt-projete}.

\bigskip

\begin{figure}    [H]  \centering
\centerline {\includegraphics[height=.49 \textwidth]{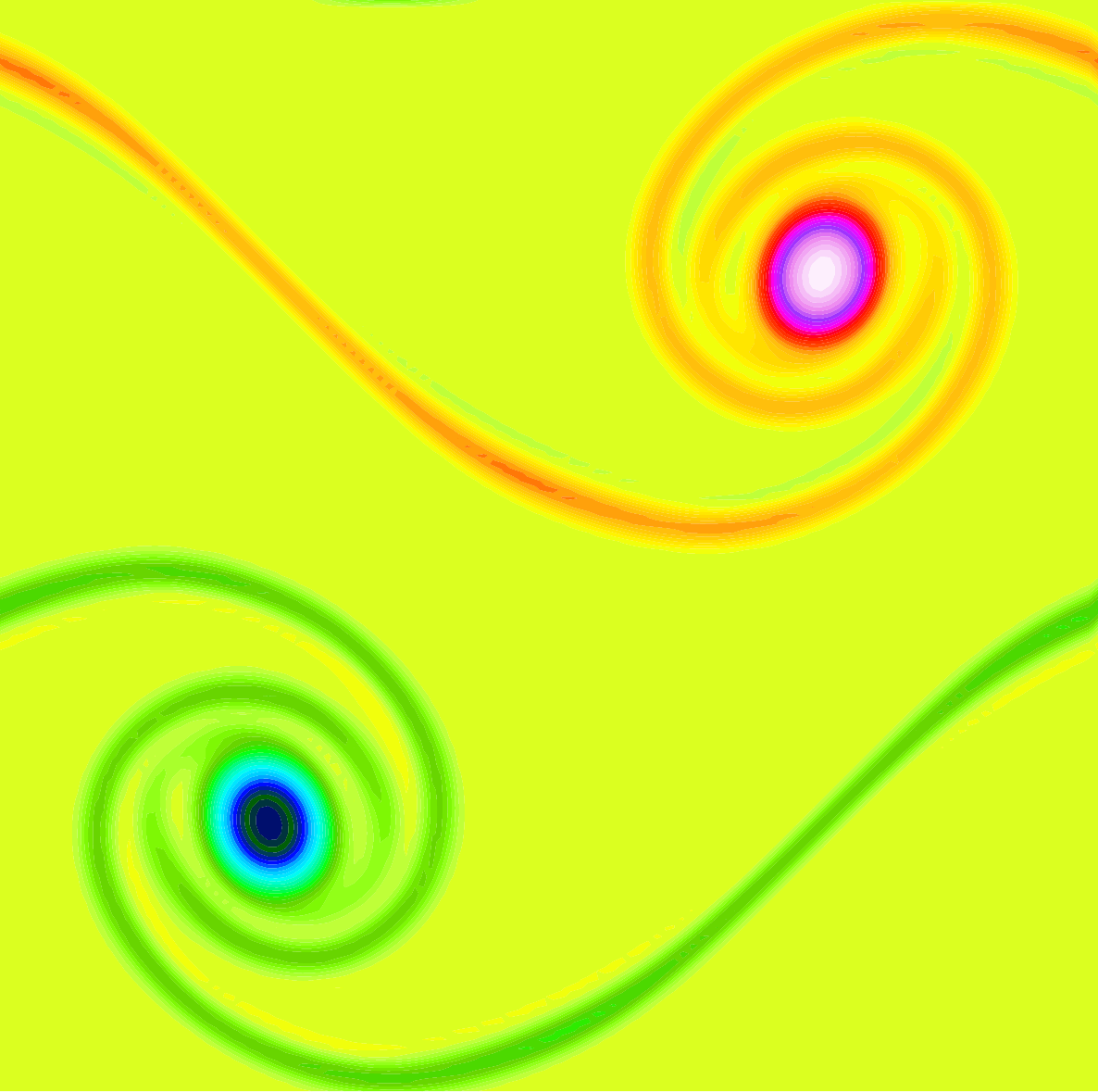}}
\caption{Minion-Brown test case \cite {MB97} for  Reynolds number $ \, Re = 10^4$,
  128 grid points and  $ \, N_T = 5541 \, $ discrete time iterations. 
  Results for the MRT scheme with projection. The bulk viscosity is reduced by using 
  the parameter $ \, s_e = 1.9999 $.   Vorticity field.}   
\label{fig-minion-vorticity} \end{figure}

It is possible to reduce the bulk viscosity for this Minion and Brown test case
with a Reynolds number $ \, Re = 10^4 $.
With the MRT with projection, we can reduce the bulk viscosity  up to $ \, \zeta =  6.51 \, 10^{-8} \, $
with  $ \, s_e  = 1.9999 $. The projected MRT lattice Boltzmann scheme remains stable. 
The Reynolds number based on this bulk viscosity
equal to  $ \,  35.5 \, 10^4 $. The results are presented in Figure \ref{fig-minion-vorticity}. 


\bigskip

\noindent
Secondly, we have considered the  lid-driven cavity initially proposed by 
Ghia {\it et al.}  \cite{GGS82}.
This test case has been simulated in the framework of lattice Boltzmann schemes
by numerous teams, including 
Guo {\it et al.}  \cite{GSW00},
Hou {\it et al.}  \cite{HZCDC95}, 
Kumar and Agrawal \cite{KA13}, 
Luo {\it et al.}~\cite{LLCPZ11}, 
Mohammedi and Reis \cite {MR17}, 
Hegele {\it et al.} \cite {HSSMPPGR18} and
Bazarin {\it et al.}  \cite {BPRH21}.

\smallskip

\noindent
The velocity $ \, U_0 \, $ on the top of the computational domain is taken equal to
$ \, U_0 = 0.01 $. It corresponds to a Mach numer of $ \, 0.0173 $.
Then the resulting flow is very close to incompressibility.
For this test case, our target Reynolds number is $ \, Re = 1.000 $.
We used a mesh with $\, N = 128 \, $ grid points as in the previous test case. 
Then from the relation (\ref{nu-sigma-dx}), we have
$ \, \sigma_\mu = 0.00384 \, $ and
$ \, s_\mu = 1.984757065735154 $.

\begin{figure}    [H]  \centering
 {\includegraphics[height=.49 \textwidth]{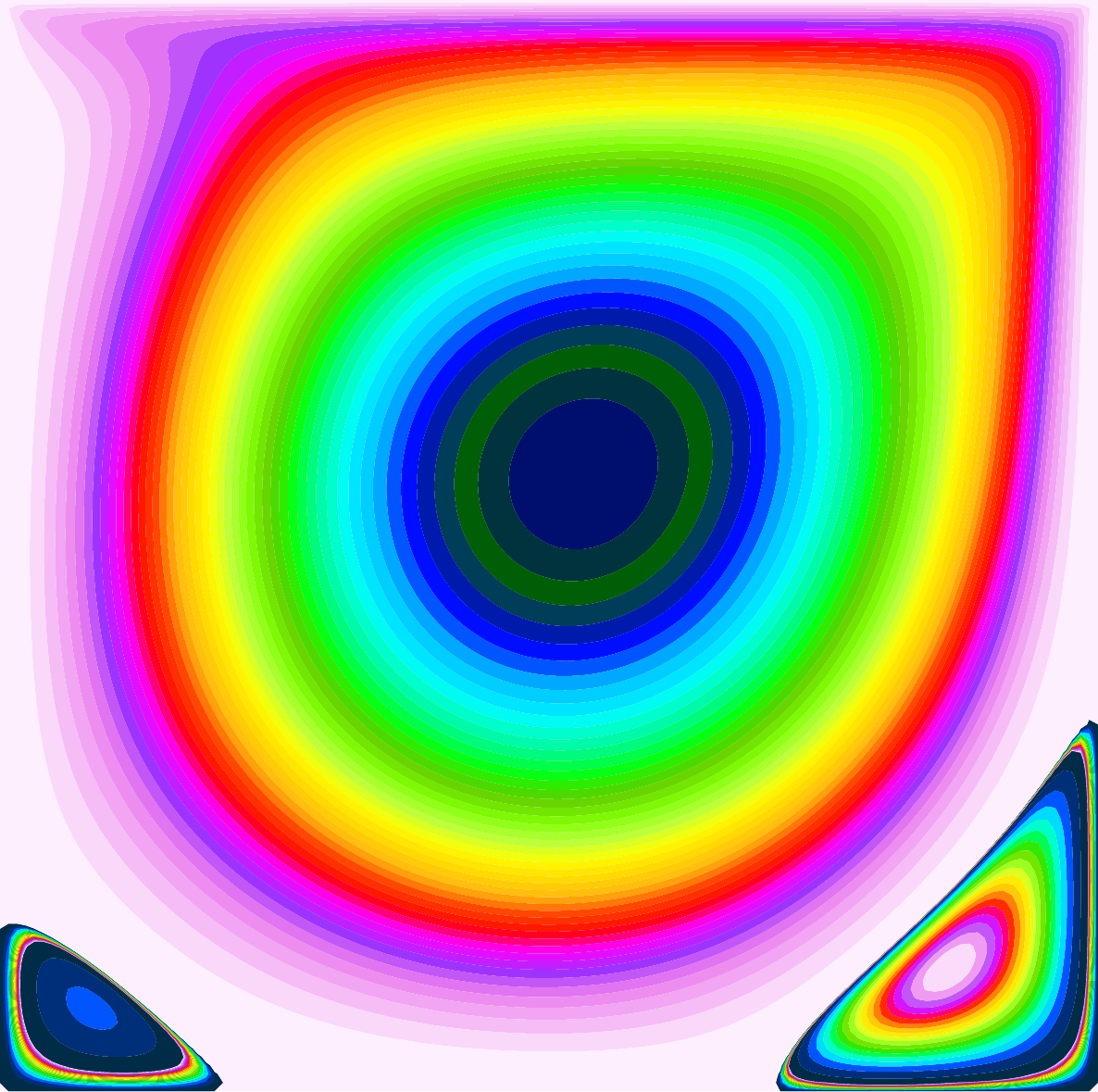}}$\,$ 
 {\includegraphics[height=.49 \textwidth]{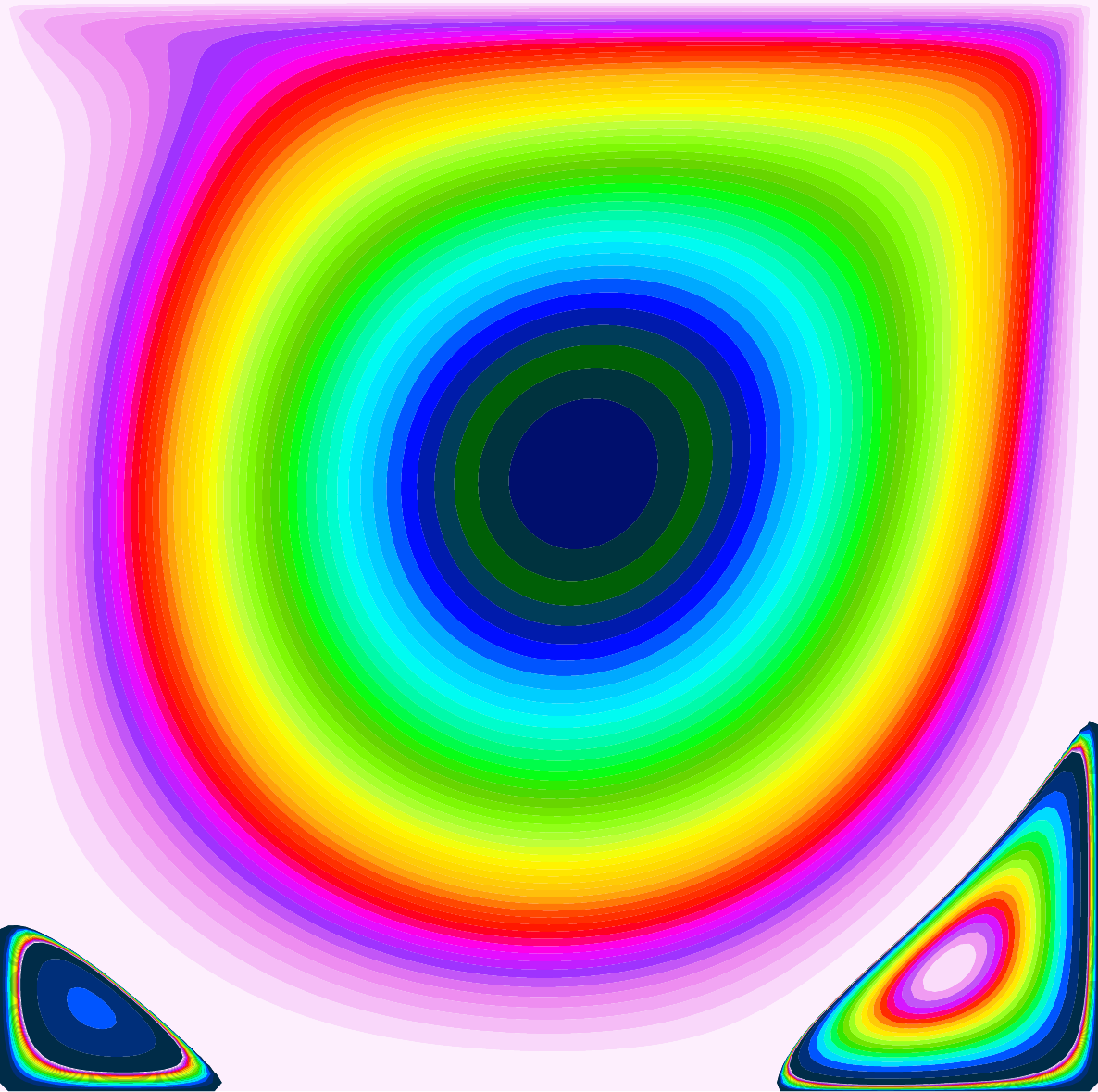}}
 \caption{Lid-driven cavity \cite{GGS82} with  $ Re = 1000 $ [$s_\mu  = 1.984757065735154$],
$  s_e = 1.7 $. Stream function for the 
   classic MRT scheme with parameters  $ s_q = s_h = 1 \, $ (left figure) and
 multiresolution relaxation times  scheme with projection (right figure).} 
\label{fig-cavity-1.7-stream-function} \end{figure}

\begin{figure}    [H]  \centering
 {\includegraphics[height=.49 \textwidth]{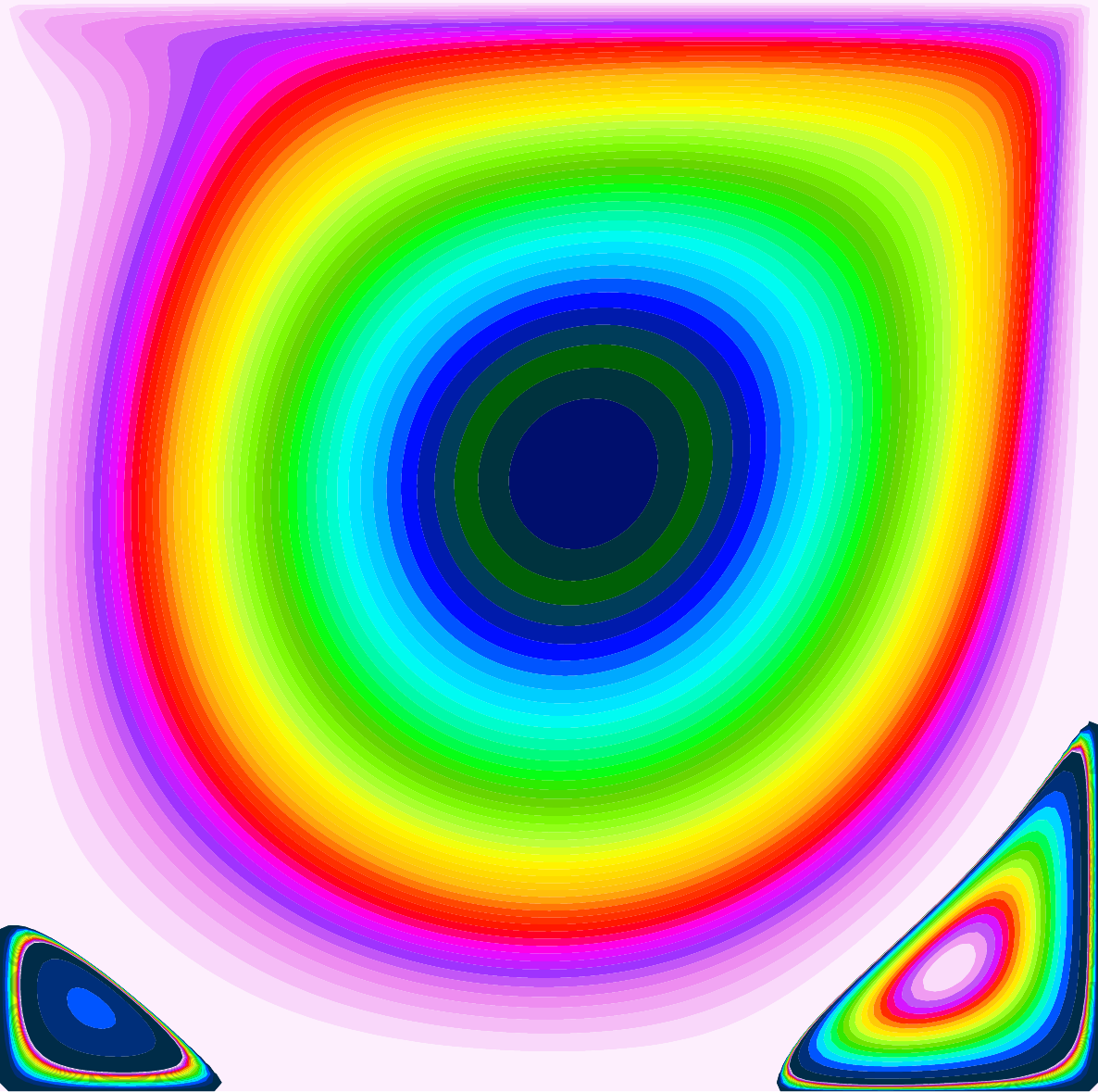}}$\,$ 
 {\includegraphics[height=.49 \textwidth]{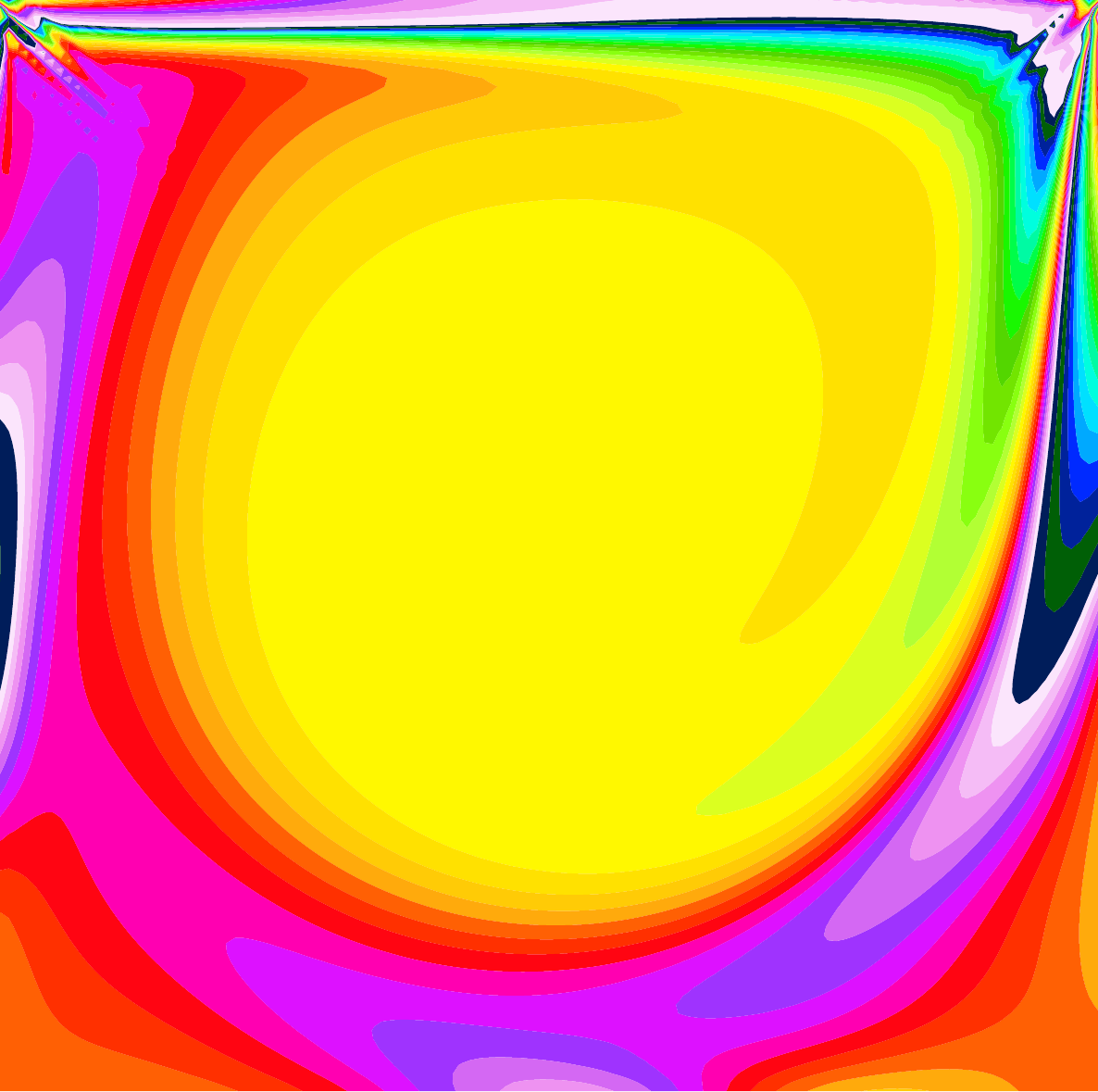}}
\caption{Lid-driven cavity \cite{GGS82} with  $ Re = 1000 $ [$s_\mu = 1.984757065735154$] and    $  s_e =  1.98 $.
Stream function  (left figure)  and vorticity  (right figure) for the  MRT scheme with projection.}
\label{fig-cavity-1.98-stream-function-vorticity} \end{figure}


\begin{figure}    [H]  \centering
  \centerline {\includegraphics[height=.55 \textwidth]{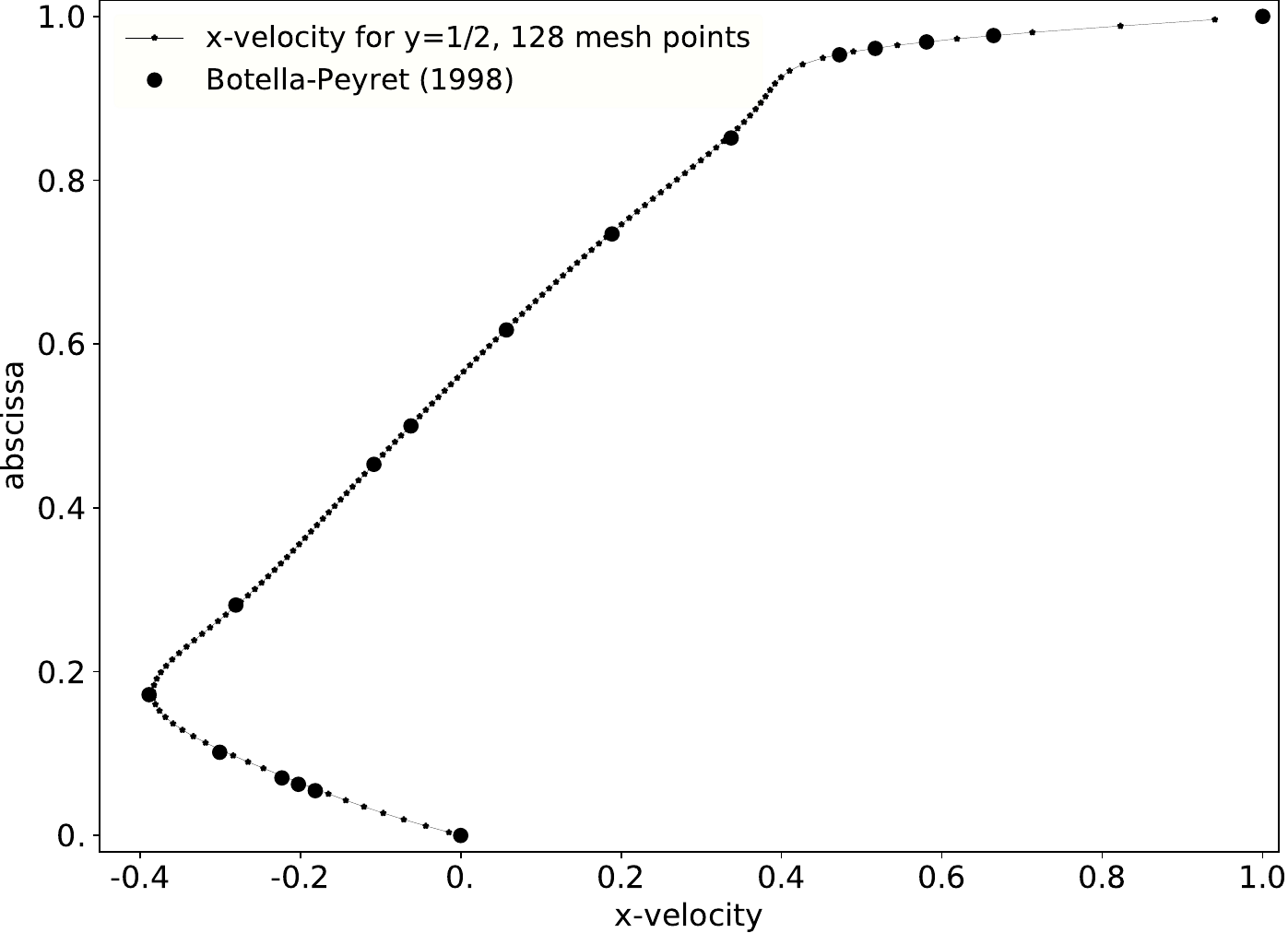}}  
\caption{Lid-driven cavity \cite{GGS82} with  $ Re = 1000 $ [$s_\mu  = 1.984757065735154$] and    $  s_e =  1.98 $.
MRT scheme with projection: $x$-component of the velocity at $ y={1\over2} \, $ and 
  comparison with Botella and Peyret \cite{BP98} results.}
\label{fig-cavity-1.98-uu} \end{figure}

\begin{figure}    [H]  \centering
  \centerline {\includegraphics[height=.55 \textwidth]{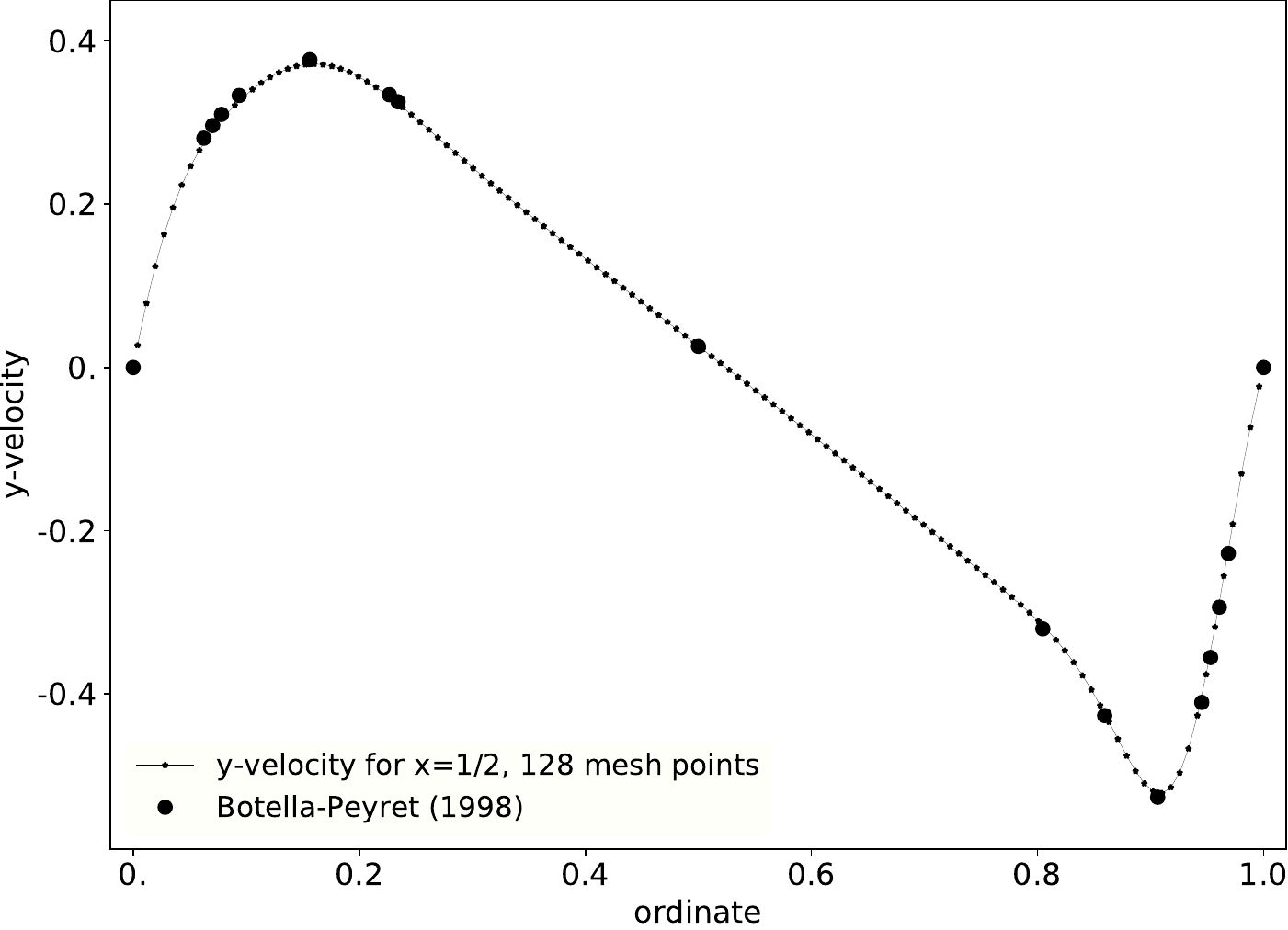}}  
\caption{Lid-driven cavity. Same test case than in Figure \ref{fig-cavity-1.98-uu};
$y$-component of the velocity at  $ x={1\over2} $.}
\label{fig-cavity-1.98-vv} \end{figure}

\bigskip   \noindent
With a relative high value of the bulk viscosity, $ \, s_e = 1.7 \, $ to fix the ideas,
it has been possible to integrate the initial multiresolution relaxation times scheme
with $ \, s_q = s_h = 1 \, $ 
up to a stationary solution. We have used  $ \, N_T = 400.000 \, $ time steps
by initializing the velocity field to zero. 
With the same parameters, the  projected multiresolution relaxation times
lattice Boltzmann scheme proposes also a stationary fluid flow.
They are compared in Figure \ref{fig-cavity-1.7-stream-function}.
Observe that taking $ \, s_e = 1.80 $, the classic MRT solver is diverging.
%
On the other hand, the projection version  gives results that make sense for fluid mechanics.
In Figure \ref{fig-cavity-1.98-stream-function-vorticity} , we show the results
for $ \,  s_e =  1.98 $, with both the current stream function and the vorticity.

\smallskip  
\noindent
In figures \ref{fig-cavity-1.98-uu} and \ref{fig-cavity-1.98-vv}, we present classical outputs for the Ghia {\it et al.}  test case:
the two components of the velocity in the middle of the flow. We compare our results with the reference
proposed by Botella and Peyret \cite{BP98} with a spectral approach. 

\noindent
Our results are quantitavely correct. We consider that these first results validate 
the  multiresolution relaxation times scheme with projection  for stationary nearly incompressible fluid flows.

\bigskip \bigskip    \noindent {\bf \large    8) \quad  Unsteady linear acoustics}


\smallskip

\noindent
With this test case, we study the two-dimensional propagation
of an initial gaussian density  profile associated with a zero velocity field.
Qualitatively, the evolution is a simple propagation of the disturbance in density
at the speed of sound $\, c_0 = {{1}\over{\sqrt{3}}} \simeq 0.5773 $. 
 It's a problem invariant to rotation around the initial center of the Gaussian.
This invariance is, of course, broken by any numerical approximation. 
With this test case, the isotropic qualities and defects of the numerical schemes  are particularly highlighted. 

\smallskip  \noindent

\begin{figure}    [H]  \centering
\centerline  {\includegraphics[width=.68 \textwidth]    {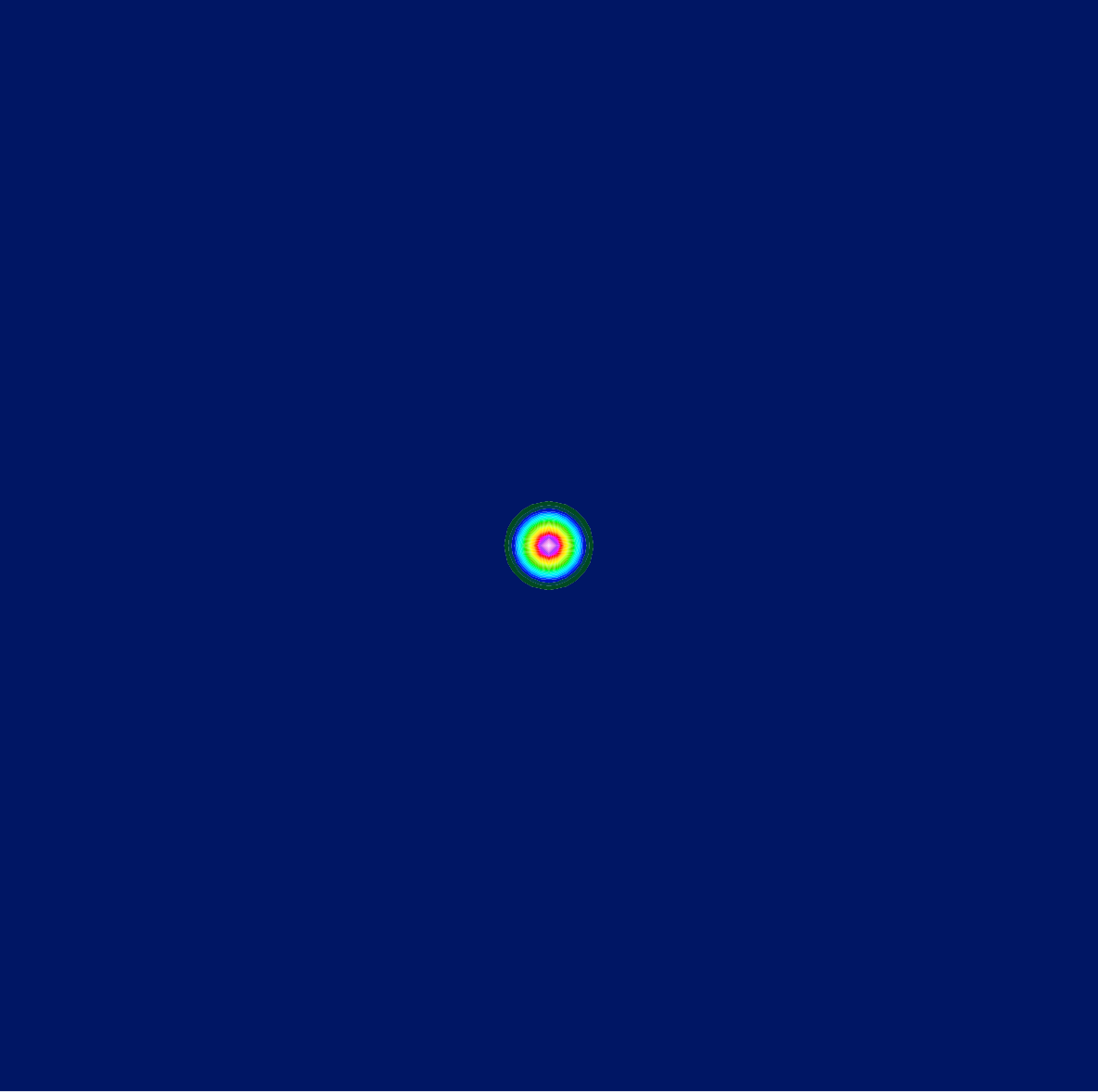}}
\caption{Unsteady acoustics: isovalues of the gaussian initial density}
\label{fig-acoustics-t0-density-isoval} \end{figure}

\begin{figure}    [H]  \centering
\centerline  {\includegraphics[width=.75 \textwidth]    {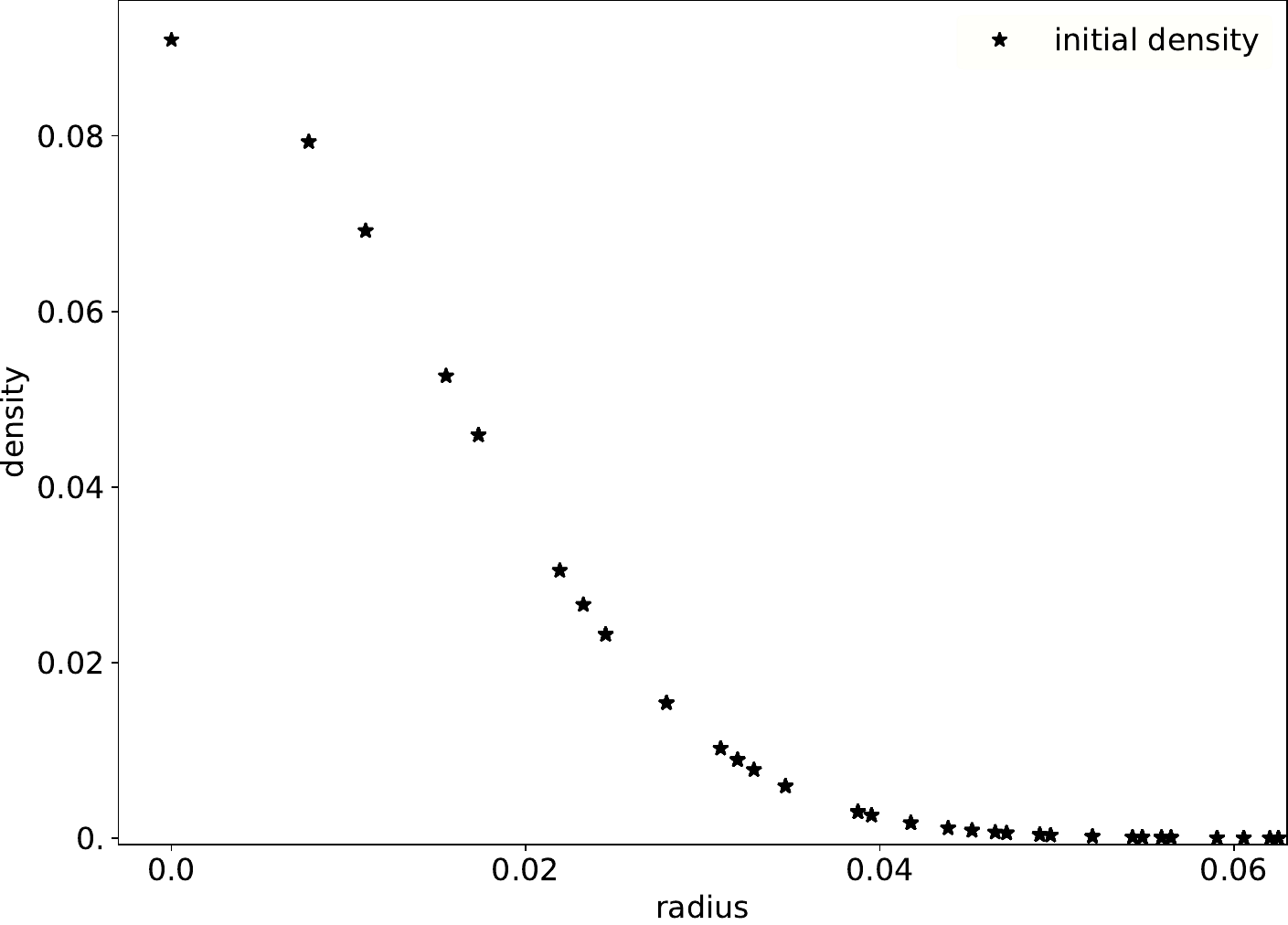}}  
\caption{Unsteady acoustics: the gaussian initial density as a function of the radius.}
\label{fig-acoustics-t0-density-rho2r} \end{figure}

\begin{figure}    [H]  \centering
\centerline  {\includegraphics[width=.68 \textwidth]    {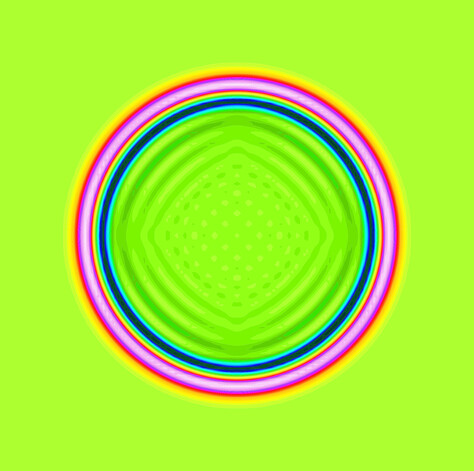}}
\caption{Acoustic with the classic MRT, $ s_\mu =1.99$,   $ s_e=1.99$, $s_q=1.9$,  $ s_h=1.54$, isovalues of the density.}
\label{fig-acoustics-MRT-se-1.99-sh-1.54-isoval} \end{figure}

\begin{figure}    [H]  \centering
    \centerline  {\includegraphics[width=.68 \textwidth]    {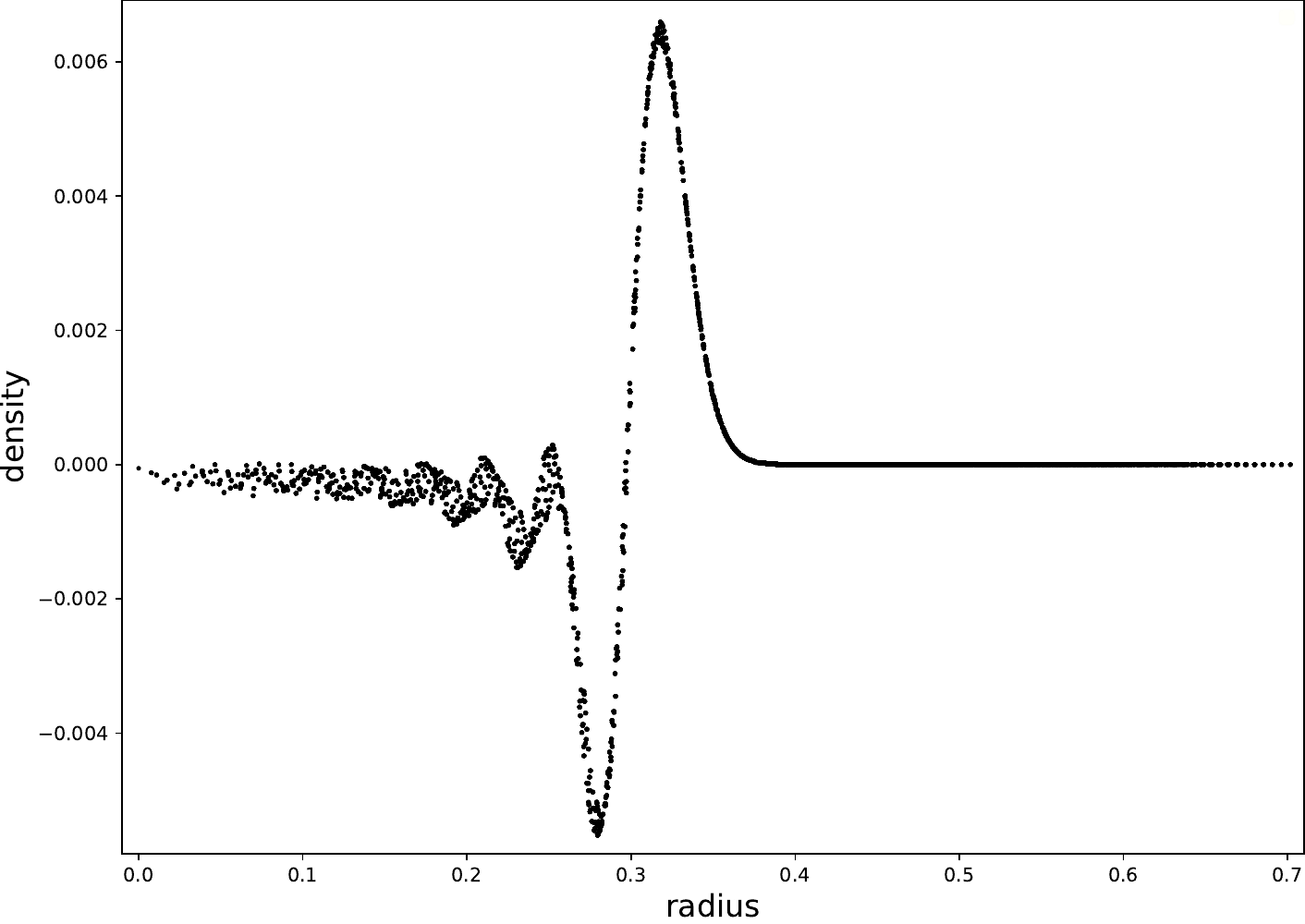}}
\caption{Unsteady acoustics with the classic  MRT scheme,
  $ s_\mu =1.99$,   $ s_e=1.99$, $s_q=1.9$,  $ s_h=1.54$, density as a function of the distance to the center.}
\label{fig-acoustics-MRT-se-1.99-sh-1.54-rho2r} \end{figure}

\begin{figure}    [H]  \centering
\centerline  {\includegraphics[width=.68 \textwidth]    {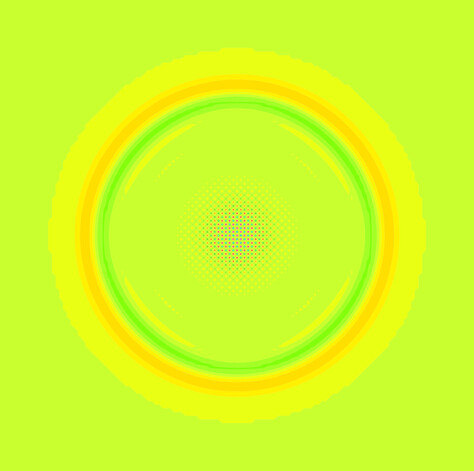}}
\caption{Unsteady acoustics with the classic, $ s_\mu =1.99 = s_e=1.99$, $s_q= s_h=1$, isovalues of the density.
Strong oscillations are clealy visible.}
\label{fig-acoustics-MRT-se-1.99-sh-1-isoval} \end{figure}

\begin{figure}    [H]  \centering
  \centerline  {\includegraphics[width=0.73 \textwidth]    {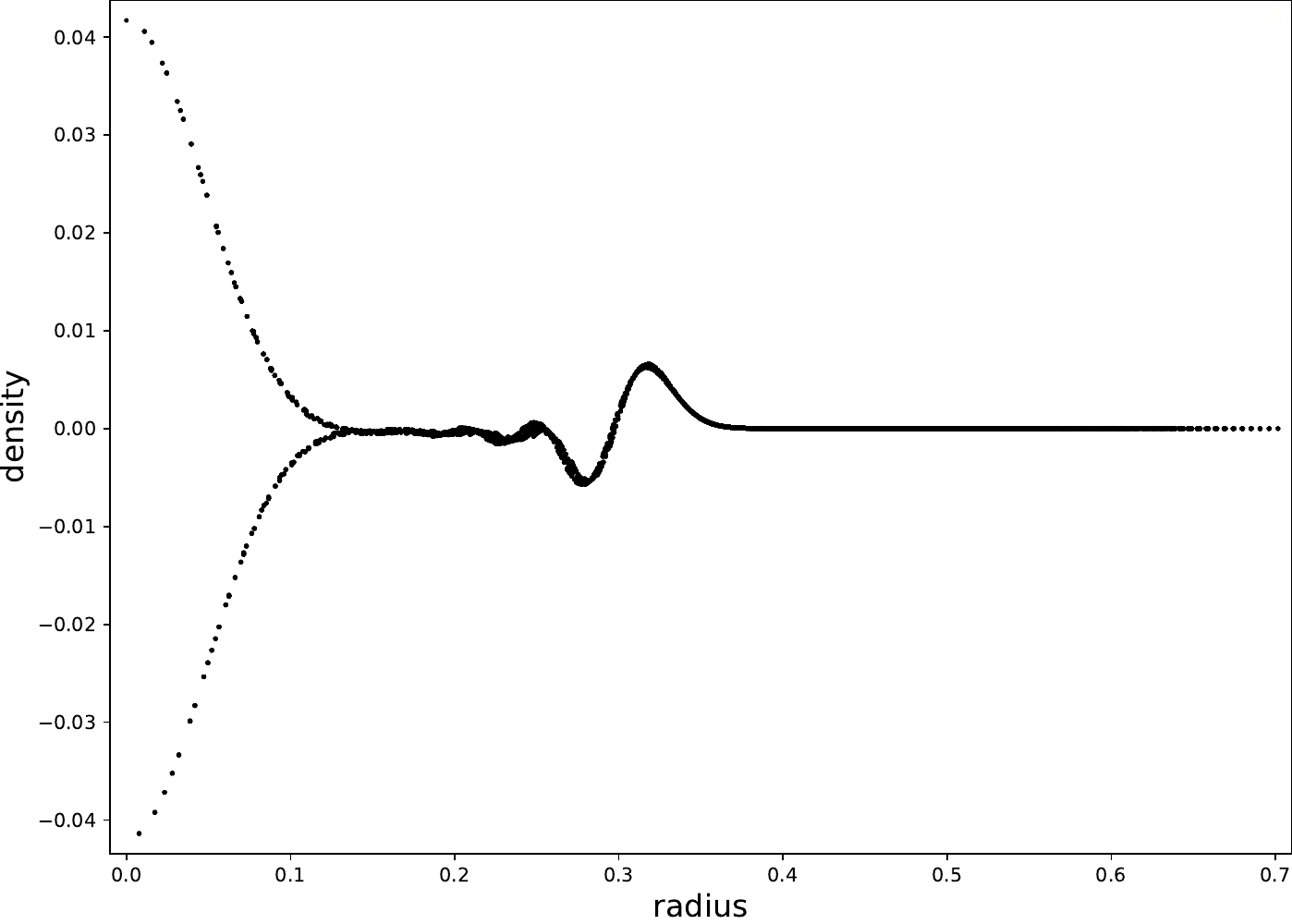}}
\vskip -.1 cm 
\caption{Unsteady acoustics: MRT, $ s_\mu =1.99 = s_e=1.99$, $s_q= s_h=1$, density as a function of the distance to the center.}
\label{fig-acoustics-MRT-se-1.99-sh-1-rho2r} \end{figure}

We took a number a meshes $ \, N = 129 \, $ in order to have 
a good location of the mesh center. 
The initial condition is a gaussian profile for the density: 
\moneqstar 
\rho = 1 +  \delta \rho \, \exp \big( -r^2 / R^2 \big)
\monendstar
with  $\, \delta \rho = 0.1 \, $ and  $\, R = 0.02 $.
This gaussian initial condition is represented in figures \ref{fig-acoustics-t0-density-isoval}
and \ref{fig-acoustics-t0-density-rho2r}.
We first use a classic MRT scheme with $ \, s_\mu  = 1.99 $.
It corresponds to a Reynolds number  $\, Re = 88927 \, $ based on  the speed of sound.
We have taken the following relaxation parameters:
\moneqstar 
 s_e =  1.99 \,,\,\,  s_q =   1.9 \,,\,\,  s_h = 1.54 \, .  
\monendstar 
The results are shown in figures \ref{fig-acoustics-MRT-se-1.99-sh-1.54-isoval} and
\ref{fig-acoustics-MRT-se-1.99-sh-1.54-rho2r}. The results are globally satisfying.
If we want to mimic the projection scheme, we change the two last relaxation parameters
for $ \, s_q = s_h = 1 $. The results are displayed in figures \ref{fig-acoustics-MRT-se-1.99-sh-1-isoval}
and \ref{fig-acoustics-MRT-se-1.99-sh-1-rho2r}. They clearly indicate than an instability is developing.

\begin{figure}    [H]  \centering
\centerline  {\includegraphics[width=.68 \textwidth]    {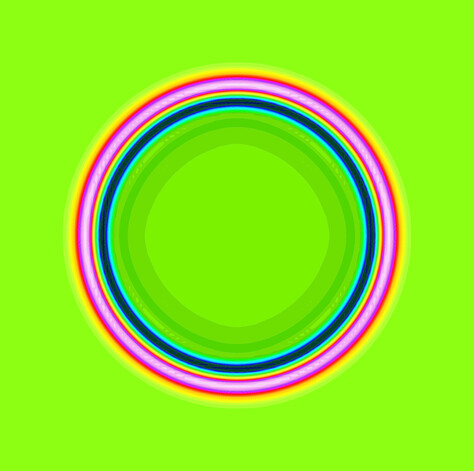}}
\caption{Unsteady acoustics, MRT with projection with parameters $ s_\mu  = s_e=1.99$, isovalues of the density.}
\label{fig-acoustics-projection-isoval} \end{figure}

\begin{figure}    [H]  \centering
  \centerline  {\includegraphics[width=0.73 \textwidth]    {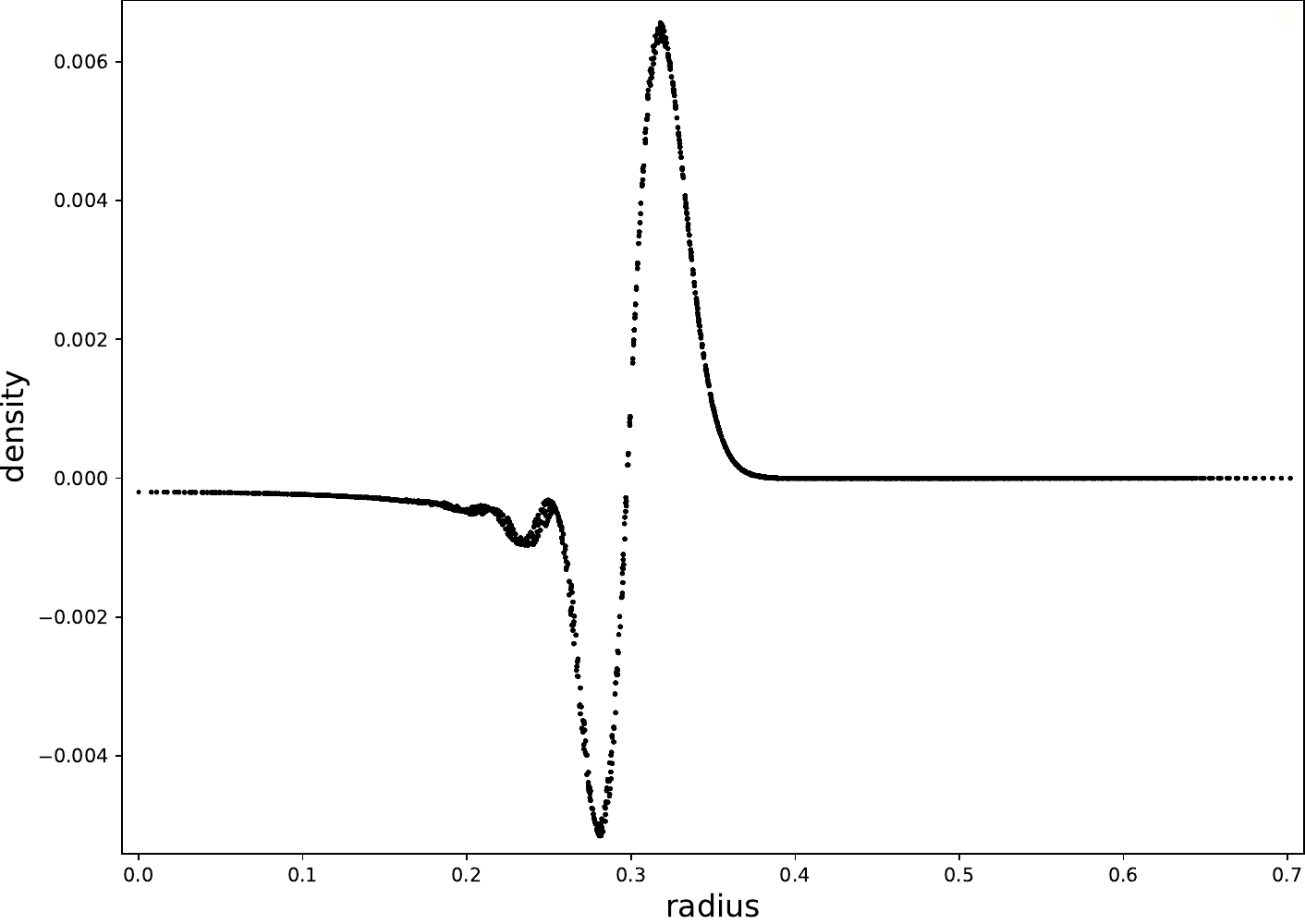}}
\caption{Unsteady acoustics, MRT with projection,  $ s_\mu  = s_e=1.99$, density as a function of the distance to the center.}
\label{fig-acoustics-projection-rho2r} \end{figure}

\smallskip  \noindent
With the projection  multiresolution relaxation times scheme for the same parameters,
{\it id est} $ \, s_\mu = s_e = 1.99 $, the results are very satisfying. 
We refer to the figures \ref{fig-acoustics-projection-isoval} and 
\ref{fig-acoustics-projection-rho2r}.
Rotation invariance is much better satisfied, as shown by the density curve
as a function of distance from the center.

\newpage 

\newpage 

\vskip -.4 cm 

\newpage 

\vskip -.4 cm

\newpage 

\vskip -.4 cm

\newpage

\begin{figure}    [H]  \centering
\centerline  {\includegraphics[width=.68 \textwidth]    {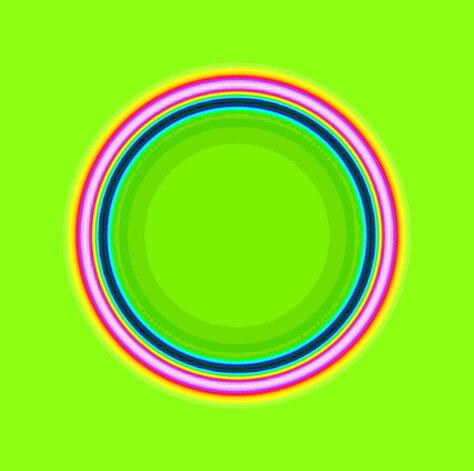}}
\caption{Unsteady acoustics: classic MRT scheme with quartic parameters,
  $ s_\mu = s_e = s_h=1.99$,  $\sigma_\mu \, \sigma_q= {1\over6}$. Isovalues of the density.}
\label{fig-acoustics-MRT-Augier-isoval} \end{figure}

\vskip -.4 cm

\begin{figure}    [H]  \centering
  \centerline  {\includegraphics[width=0.73 \textwidth]    {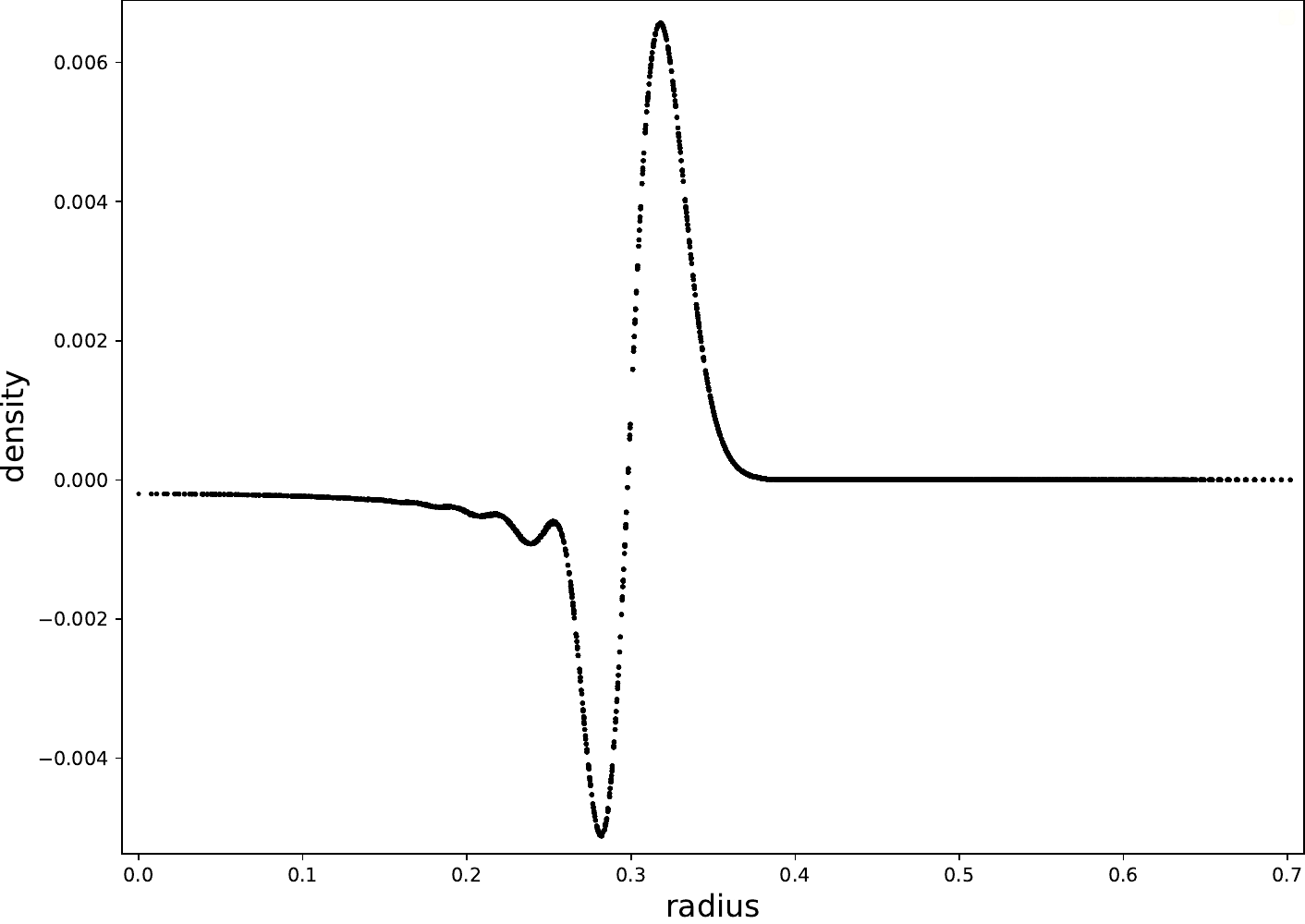}}
\vskip -.1 cm 
\caption{Unsteady acoustics: classic MRT scheme,  $ s_\mu = s_e = s_h=1.99$,   $ \sigma_\mu  \, \sigma_q= {1\over6}$,
  density as a function of the distance to the center.}
\label{fig-acoustics-MRT-Augier-rho2r} \end{figure}

\smallskip  \noindent
Last but not least, Augier {\it et al.} have studied in \cite{ACS93}
the possibility of rotation-invariant MRT lattice Boltzmann schemes up to fourth order.
In the D2Q9 case, the ``quartic'' relaxation parameters take the form
\moneqstar
\sigma_\mu  = \sigma_e = \sigma_h \,,\,\, \sigma_\mu  \, \sigma_q = {1\over6} \, .
\monendstar
In our case, $ \, s_\mu  = s_e = s_h = 1.99\, $ and the very unusual value $ \, s_q = 0.01496259351620921 $.
The results are presented in figures \ref{fig-acoustics-MRT-Augier-isoval} and 
 \ref{fig-acoustics-MRT-Augier-rho2r}. 
 They are very good quality.
 
\smallskip
As a conclusion of this section, the projected MRT  scheme has 
a very good ability to give correct results with good stability.
The initial version of the MRT scheme uses more parameters, is more fragile
from a stability point of view, but in some cases produces better quality results.

\bigskip \bigskip    \noindent {\bf \large    9) \quad  Conclusion}

\smallskip  \noindent
We have extended the lattice Boltzmann introduced by Malaspinas \cite{Ma15} 
and developed also by Mattila {\it et al.} \cite {MPH17}.
Our approach is no longer based on Hermite polynomials of the moments
but on the spectific structure (\ref{Lambda-d2q9-blocs}) of the advection
operator in the basis of moments.
One advantage of this projection scheme is the reduction in the number of relaxation parameters.
%
Our analysis at second order accuracy shows that the equivalent partial differential equations
are not modified by the  addition of a projection step in the algorithm.
The first numerical tests are satisfying and highlights a gain in stability
for the parameters associated to bulk viscosity.

\smallskip  \noindent
The natural next step is to study lattice Boltzmann schemes in three spatial dimensions.
Our results with Pierre  Lallemand \cite{DL23} indicate a favorable
block structure for the advection operator in the basis of moments.


\bigskip \bigskip    \noindent {\bf \large    10) \quad  Annex: equivalent equations of the projection scheme}

%
\smallskip \noindent 
We recall the Proposition 1 statement. 
We suppose that the advection matrix in the basis of moments
satisfies the condition (\ref{Lambda-d2q9-blocs}).
We suppose morever that after the decompostion (\ref{moments-d2q9-3-blocs})
of the moments into conserved variables $\, W $,  eulerian moments$ \, Y_e \, $
and viscous moments $ \, Y_w $, the equilibrium values satisfy the relation 
(\ref{lien-Phiv-Phie}).
Then the MRT scheme with projection, defined by the relations (\ref{projecteur-d2q9})
to   (\ref{propagation}), satisfies at order two the asymptotic  relations
(\ref{equivalent-edp-ordre-2-schema-avec -projection}): 
\moneqstar 
\left\{ \begin{array} {l}
 \partial_t W + \Gamma_1 + \Delta t \, \Gamma_2 = {\rm O}(\Delta t^2)  \\ 
  \Gamma_1 = A \, W +  B_e  \, \Phi_e   \\
  \Psi_e  =  \dd \Phi_e (W) .  \Gamma_1 - ( C_e  \, W +  D_{ev}  \, \Phi_v (W))  \\
  \Gamma_2 =   B_e  \, \Sigma_e \Psi_e \, . 
\end{array} \right. \monendstar

\bigskip

We begin by an analysis at order zero. 
We still have the relation (\ref{m-exponentielle-m-star}) 
\moneqstar
m  (x, t + \Delta t) =  {\rm exp} ( - \Delta t\, \Lambda ) \,\,  m^*(x ,\, t) \, .
\monendstar 
Moreover, the  moments $ \, m^* \, $  after relaxation satisfy now:
\moneqstar
m^* = ( W ,\, Y_e^* ,\, K \, W + L \,  Y_e^* )^{\rm t} \, .
\monendstar
In consequence, we have  $\, m + {\rm O}(\Delta t) =  m^* + {\rm O}(\Delta x) \, $
and in particular
\moneqstar
Y_e - Y_e^* \equiv S_e \, (Y_e - \Phi_e ) = {\rm O}(\Delta t) \, .
\monendstar
Then  $ \, Y_e = \Phi_e +  {\rm O}(\Delta t)  \, $
when $ \, S_e \, $ is fixed and is invertible. Then we have also 
\moneqstar
Y_e^* = \Phi_e +  {\rm O}(\Delta t)  .
\monendstar
For the third component, we have due to the relation (\ref{Yv-star-d2q9}), 
\moneqstar
 Y_v^* \equiv   K \, W + L \,  Y_e^* 
  =  K \, W + L \, \Phi_e  +  {\rm O}(\Delta t)   =    \Phi_v  +  {\rm O}(\Delta t) \, .
\monendstar
We have also 
$\, (P m)_v + {\rm O}(\Delta t) =  Y_v^* + {\rm O}(\Delta x) \, $
and $ \,  (P m)_v  = \Phi_v  + {\rm O}(\Delta x) $.
Then by combining the three components, we have at order zero 
\moneqstar
m = m^{\rm eq} + {\rm O}(\Delta t)
\monendstar
and 
\moneqstar
m^* = m^{\rm eq} + {\rm O}(\Delta t)  \, .
\monendstar

\bigskip

We consider now the  analysis at order one. We expand the     relation (\ref{m-exponentielle-m-star}) 
at first order. For the first component $\, W $, we have 

\smallskip  \noindent 
 $ \, W + \Delta t \, \partial_t W +  {\rm O}(\Delta t^2) =
W -  \Delta t \,  (A \, W + B \, Y^* ) +   {\rm O}(\Delta t^2)  $

\smallskip  \qquad \qquad  \qquad \qquad  \qquad 
$ \,\,\,  = W -  \Delta t \,  (A \, W + B_e \, Y_e^* ) +   {\rm O}(\Delta t^2)  $.

\smallskip  \noindent
Then we have 
$ \, \partial_t W +  (A \, W + B_e \, Y_e^* ) =  {\rm O}(\Delta t) \, $
and due to the identity 
$ \,  Y_e^* = \Phi_e +  {\rm O}(\Delta t) $, 
the relation
    $  \, \partial_{t} W +  {\boldsymbol A}  \, W + B_e   \, \Phi_e(W) =  {\rm O}(\Delta t) \, $
is  still true. In conclusion, we have at first order 
 $ \,  \Gamma_1 = A   \, W +  B_e  \, \Phi_e(W) $.

\bigskip

We consider again now the analysis at order one.
We pay attention to the fact that the devil  in the details!
We expand the relation relation (\ref{m-exponentielle-m-star}) 
at first order and we focus on the second component. We have 
\moneqstar
Y_e + \Delta t \, \partial_t Y_e +  {\rm O}(\Delta t^2) =
Y_e^*  - \Delta t \, (C_e \, W + D_{ev} \, Y_v^* ) +   {\rm O}(\Delta t^2)
\monendstar
and we remark that
$ \, \partial_t Y_e = \dd \Phi_e . \partial_t W  + {\rm O}(\Delta t)  =- \dd \Phi_e . \Gamma_1 + {\rm O}(\Delta t) $.
Than 
\moneqstar
Y_e  -  \Delta t \,  \dd  \Phi_e . \Gamma_1  +  {\rm O}(\Delta t^2) =
Y_e^*  - \Delta t \, (C_e \, W + D_{ev} \,  \Phi_v   ) +   {\rm O}(\Delta t^2) \, .
\monendstar
We deduce that $\,\,  S_e \, ( Y_e - \Phi_e ) \equiv  Y_e - Y_e^*  =
\Delta t \, [  \dd  \Phi_e . \Gamma_1 - (C_e \, W + D_{ev} \, \Phi_v ) ] +   {\rm O}(\Delta t^2) $.
With   $ \,  \Psi_e  =  \dd \Phi_e (W) .  \Gamma_1 - ( C_e  \, W +  D_{ev}  \, \Phi_v (W)) $,
we can expand this family of non-conserved moments at first order: 
$\, Y_e = \Phi_e  - \Delta t \, S_e^{-1} \,  \Psi_e + {\rm O}(\Delta t^2) $.
Then the relaxation scheme
\moneqstar
Y_e^* = ({\rm I} - S_e) \, Y_e + S_e \,  \Phi_e  
\monendstar
implies

\smallskip \noindent 
$  Y_e^* = Y_e - S_e \, ( Y_e - \Phi_e ) \, = \Phi_e  - \Delta t \, S_e^{-1} \,  \Psi_e
+ \Delta t \, \Psi_e + {\rm O}(\Delta t^2) $

\smallskip
\qquad \qquad \qquad \qquad \qquad   
$ \,\,\,  = \Phi_e + \Delta t \, \big( {\rm I} - S_e^{-1} \big) \, \Psi_e + {\rm O}(\Delta t^2) $. 

\smallskip
We introduce the ``reduced H\'enon matrix'' 
\moneqstar
\Sigma_e \equiv  S_e^{-1} - {1\over2} \, {\rm I}  \, .
\monendstar
Then we have the following 
expansions of the non-conserved moments at first order 
\moneqstar 
\left\{ \begin{array} {l}
  Y_e = \Phi_e  + \Delta t \, \big( \Sigma_e  +  {1\over2} {\rm I} \big) \, \Psi_e + {\rm O}(\Delta t^2)  \\
  Y_e^*  = \Phi_e  + \Delta t \, \big( \Sigma_e  -  {1\over2} {\rm I} \big) \, \Psi_e + {\rm O}(\Delta t^2)  .
\end{array} \right. \monendstar

\bigskip

For the analysis at order two,
we need to calculate the value of the  $ \, \Lambda^2 \, $   matrix.
We have

\smallskip 
$  \Lambda^2 = \begin{pmatrix} A   &  B_e &   0 \\
C_e  & 0 &  D_{ev}  \\ 0 &  D_{ve}  &  D_{vv}  \end  {pmatrix}  \,  \begin{pmatrix} A   &  B_e &   0 \\
  C_e  & 0 &  D_{ev}  \\ 0 &  D_{ve}  &  D_{vv}  \end  {pmatrix} $

\smallskip  \quad 
$ \, =  \begin{pmatrix} A^2 +  B_e \, C_e  & A \, B_e & B_e \, D_{ev}  \\
  C_e \, A & C_e \, B_e + D_{ev}  \, D_{ve}  &  D_{ev}  \,  D_{vv} \\
  D_{ve} \,  C_e  & D_{vv} \, D_{ve}  &  D_{ve} \,   D_{ev}  +   D_{vv}^2 \end  {pmatrix} $

\smallskip \noindent 
because 
$\, A_2 = A^2 + B_e \, C_e  \, $
and  $\, B_2 =   \big( A \, B_e  \quad   B_e \, D_{ev}  \big)   $.


\smallskip \noindent
For the analysis at order two,
we  expand the relation  (\ref{m-exponentielle-m-star})
at second order for the first component $\, W $. We obtain 

 \smallskip 
 $ \, W + \Delta t \, \partial_t W +  {1\over2}  \Delta t^2 \,  \partial_t^2 W  +  {\rm O}(\Delta t^3) $

 \smallskip \qquad 
 $ =  W -  \Delta t \,  (A \, W + B \, Y^* ) +   {1\over2}  \Delta t^2 \, (A_2 \, W + B_2 \, Y^* ) +  {\rm O}(\Delta t^3)  $

 \smallskip \qquad 
 $  = W -  \Delta t \,  (A \, W + B_e \,  Y_e^*  ) $$  +   {1\over2}  \Delta t^2 \, \big(  (A^2 +  B_e \, C_e) \, W + 
 A \, B_e  \, Y_e^* + B_e  \, D_{ev}   \, Y_v^* \big) +   {\rm O}(\Delta t^3)  $.

 \smallskip 
 We  observe that
 
 \smallskip \noindent 
$  \partial_t^2 W = -\partial_t \big( A \, W + B_e \, \Phi_e \big) +  {\rm O}(\Delta t)  $

 \smallskip \qquad  
 $ \,\,  = - \big( A + B_e \, \dd \Phi_e \big) . \partial_t W +  {\rm O}(\Delta t)  $
 $ =  ( A + B_e \, \dd \Phi_e ) . \Gamma_1  +  {\rm O}(\Delta t)  $. 

 \smallskip 
Then 
 
\smallskip \noindent 
$  \partial_t W = - {{\Delta t}\over2} ( A + B_e \, \dd \Phi_e ) . \Gamma_1   - A \, W $ 
 $ - B_e  \big( \Phi_e  + \Delta t \, \big( \Sigma_e  -  {1\over2} {\rm I} \big) \, \Psi_e  \big) $ 

 \smallskip \qquad  \qquad 
$  + {{\Delta t}\over2} \big[ (A^2 +  B_e \, C_e) \, W + 
 A \, B_e  \, \Phi_e  + B_e  \, D_{ev}   \, \Phi_v    \big] +  {\rm O}(\Delta t^2)  $

\smallskip  \qquad 
$ =  -( A \, W +  B_e \, \Phi_e ) - \Delta t \,  B_e \, \Sigma_e \Psi_e  +  \Delta t  \big[
 - {1\over2} ( A + B_e \, \dd \Phi_e ) . \Gamma_1 $ 

 \smallskip \qquad  \qquad 
 $ + {1\over2} B_e \,( \dd \Phi_e .  \Gamma_1 - C_e \, W - D_{ev} \, \Phi_v )
 + {1\over2} A \, (A \, W + B_e  \, \Phi_e) $

 \smallskip \qquad  \qquad 
 $ +   {1\over2}  B_e \, (C_e  \, W  +  D_{ev}   \, \Phi_v )  \big]  +  {\rm O}(\Delta t^2)  $

\smallskip  \qquad 
$ =  -( A \, W +  B_e \, \Phi_e ) - \Delta t \,  B_e \, \Sigma_e \Psi_e  +   {{\Delta t}\over2} \big[
  -  A\, \Gamma_1     -   B_e \, \dd \Phi_e  . \Gamma_1 $

 \smallskip \qquad  \qquad 
  $ +   B_e \, \dd \Phi_e  . \Gamma_1  - B_e \, ( C_e \, W + D_{ev} \, \Phi_v ) +   A\, \Gamma_1  $
$ + B_e \, ( C_e \, W + D_{ev} \, \Phi_v )  \big] +  {\rm O}(\Delta t^2)  $

 \smallskip \qquad  
$ = -( A \, W +  B_e \, \Phi_e ) - \Delta t \,  B_e \, \Sigma_e \Psi_e +  {\rm O}(\Delta t^2)  $.

\smallskip  \smallskip \smallskip 
We have finally 
\moneqstar
\partial_t W + \Gamma_1 + \Delta t \, \Gamma_2 = {\rm O}(\Delta t^2)
\monendstar
with $ \,\,   \Gamma_1 = A \, W + B_e \, \Phi_e  \,  $ and  $\,  \Gamma_2 =  B_e \, \Sigma_e \Psi_e  $.
The relations
(\ref{equivalent-edp-ordre-2-schema-avec -projection})
are established. \hfill $\square$ 


%
\bigskip   \bigskip
\noindent {\bf  \large  Acknowledgments }

\smallskip
This work was initiated in summer 2018 during FD's stay in Curitiba
following PP's invitation to the  Mechanical Engineering Department
of the Catholic University of Parana in Curitiba (Paran\'a, Brazil).


\newpage
\bigskip \bigskip      \noindent {\bf  \large  References }


\end{document}